\documentclass[aps,preprint,nofootinbib]{revtex4}
\usepackage{amsmath}
\usepackage{graphicx}

\makeatletter
\allowdisplaybreaks[2]

\begin{document}
\newcommand{\pslash}[1]{\not{\!#1}}

\title{Demonstration of One Cutoff Phase Space Slicing Method:\\
Next-to-Leading Order QCD Corrections to the $tW$ Associated Production
in Hadron Collision\\
\vspace*{1cm}}

\author{Qing-Hong Cao}

\email{qcao@ucr.edu}

\address{Department of Physics and Astronomy, University of California at
Riverside, Riverside, CA 92531, USA\vspace{1in}
}

\begin{abstract}
We present a detailed calculation of next-to-leading order QCD corrections
to the $Wt$ associated production using the one cutoff phase space
slicing method. Such QCD corrections have been calculated independently
by two groups already, however, a number of differences were found.
It is desirable to have a third party calculation to make a crossing
check. In this note, we present our complete results of the virtual
corrections which are not shown in the literature so far. The numerical
comparison will be presented in the forthcoming paper. As a demonstration
of the one cutoff phase space slicing method, we also show in details
how to organize the color ordered amplitudes and how to slice the
soft and collinear phase space regions. 
\end{abstract}
\maketitle

\section{introduction}

Single top production at the hadron collider has been extensively
studied in the literature \citep{Tait:1997fe,Dawson:1984gx,Willenbrock:1986cr,Yuan:1989tc,Ladinsky:1990ut,Cortese:1991fw,Ellis:1992yw,Carlson:1993dt,Bordes:1994ki,Stelzer:1995mi,Heinson:1996zm,Smith:1996ij,Mrenna:1997wp,Stelzer:1997ns,Moretti:1997ng,Stelzer:1998ni,Belyaev:1998dn,Tait:1999cf,Belyaev:2000me,Tait:2000sh,Beccaria:2006ir,Beccaria:2007tc,Kidonakis:2006bu,Kidonakis:2007ej,Cao:2007ea,Gerber:2007xk}.
There are three separate single top quark production processes of
interest at the hadron collider, which may be characterized by the
virtuality of the $W$ boson (of four momentum $q$) in the processes.
The s-channel process $q\bar{q}^{\prime}\rightarrow W^{*}\rightarrow t\bar{b}$
via a virtual s-channel $W$ boson involves a timelike $W$ boson,
$q^{2}>(m_{t}+m_{b})^{2}$, the t-channel process $qb\rightarrow q't$
(including $\bar{q}'b\rightarrow\bar{qt}$, also referred as $W$-gluon
fusion) involves a spacelike $W$ boson, $q^{2}<0$, and the $tW$
associated production process $bg\rightarrow tW^{-}$ involves an
on-shell $W$ boson, $q^{2}=m_{W}^{2}$. Therefore, these three single
top quark production mechanisms probe the charged-current interaction
in different $q^{2}$ regions and are thus complementary to each other.
To improve the theory prediction on the single top production rate,
the next-to-leading order (NLO) corrections, at the order of $\alpha_{s}$,
for these three channels has been carried out in Refs.\ \citep{Harris:2002md,Zhu:2002uj,Campbell:2004ch,Cao:2004ap,Cao:2004ky,Sullivan:2004ie,Cao:2004Pri2,Frixione:2005vw,Campbell:2005bb}. 

The NLO QCD corrections to the $tW$ associated production have been
calculated independently by two groups\ \citep{Zhu:2002uj,Campbell:2005bb},
but a number of differences were found. It is desirable to have a
third party calculation to make a crossing check. Furthermore, there
is no analytic result available in the literature so far. In this
note we present a detailed calculation of the NLO QCD corrections
to the $tW$ associated production using one cutoff phase space slicing
method. The rest of this paper is organized as follows. In Sec. II,
we briefly review the method of our calculation. In Sec. III, we present
the Born level helicity amplitudes of the $tW$ production. In Sec.
IV we present the NLO virtual corrections. In Sec. V we present the
calculation of the soft and collinear singularities of the real emission
correction using the one cutoff phase space slicing method.

\section{Short review of the phase slicing method}

The construction of a flexible event generator requires the generation
of partonic final states with a minimal amount of implicit phase space
integration. At the leading order this is trivial, but in the calculation
of NLO QCD corrections, one generally encounters both ultraviolet
(UV) and infrared (IR) (soft and collinear) divergences. The former
divergences can be removed by proper renormalization of couplings
and wave functions. In order to handle the latter divergences, one
has to consider both virtual and real corrections and carefully handle
the cancellation of divergences between the soft and collinear contributions
and the virtual corrections. The soft divergences will cancel according
to the Kinoshita-Lee-Nauenberg (KLN) theorem \citep{Kinoshita:1962ur,Lee:1964is},
but some collinear divergences remain uncancelled. In the case of
considering the initial state partons, one needs to absorb additional
collinear divergences to define the NLO parton distribution function
of the initial state partons. After that, all the infrared-safe observables
will be free of any singularities. To calculate the inclusive production
rate, one can use dimensional regularization to regularize divergences
and adopt the modified minimal subtraction ($\overline{{\rm MS}}$)
factorization scheme to obtain the total rate. However, owing to the
complicated phase space for multi-parton configurations, analytic
calculations are in practice impossible for all but the simplest quantities.
During the last few years, effective numerical computational techniques
have been developed to calculate the fully differential cross section
to NLO and above. There are, broadly speaking, two types of algorithm
used for NLO calculations, differing in how they approximate the phase
space and matrix elements in the neighborhood of divergent regions:

\begin{enumerate}
\item The phase space slicing (PSS) method is based on approximating the
matrix elements and the phase space integration measure in boundary
regions of phase space so integration may be carried out analytically~\citep{Gutbrod:1983qa,Baer:1989jg,Aversa:1990uv,Giele:1991vf,Giele:1993dj,Keller:1998tf}. 
\item The subtraction method is based on adding and subtracting counterterms
designed to approximate the real emission amplitudes in the phase
space boundary regions on the one hand, and to be integrable with
respect to the momentum of an unresolved parton on the other~\citep{Ellis:1980wv,Ellis:1989vm,Bethke:1991wk,Mangano:1991jk,Frixione:1995ms,Catani:1996jh,Catani:1996vz}.
\end{enumerate}
The phase space slicing method makes use of a combination of analytic
and Monte Carlo integration methods, which has many advantages over
a purely analytic calculation. The Monte Carlo approach allows one
to calculate any number of observables simultaneously by simply histogramming
the appropriate quantities. Furthermore, it is easy to tailor the
Monte Carlo calculation to different experimental conditions, for
example, detector acceptances, experimental cuts, and jet definitions.
Also, with the Monte Carlo approach one can study the dependence of
the cross section on the choice of scale and the size of higher order
quantum corrections in different regions of phase space. The basic
challenge is to design a program which retains the versatility inherent
in a Monte Carlo approach while ensuring that all of the required
cancellations of singularities still take place. 

In this study, we use the phase space slicing method with one cutoff
scale for which the universal crossing functions have been derived
in Refs.~\citep{Giele:1991vf,Giele:1993dj,Keller:1998tf}. The advantage
of this method is that, after calculating the effective matrix elements
with all the partons in the final state, we can use the generalized
crossing property of the NLO matrix elements to calculate the corresponding
matrix elements numerically without requiring any further effort.
The validity of this method is due to the property that both the phase
space and matrix element for the initial and final state collinear
radiation processes can be simultaneously factorized. Below, we briefly
review the general formalism for the NLO calculation in PSS method
with one cutoff scale.

The phase space slicing method with one cutoff scale introduces an
unphysical parameter $s_{{\rm min}}$ to separate the real emission
correction phase space into two regions: 

\begin{enumerate}
\item the resolved region in which the amplitude has no divergences and
can be integrated numerically by Monte Carlo method; 
\item unresolved region in which the amplitude contains all the soft and
collinear divergences and can be integrated out analytically. 
\end{enumerate}
It should be emphasized that the notion of resolved/unresolved partons
is unrelated to the physical jet resolution criterium or to any other
relevant physical scale. In the massless case, a convenient definition
of the resolved region is given by the requirement\begin{equation}
s_{ij}=\left(p_{i}+p_{j}\right)^{2}>s_{min},\label{eq:condition}\end{equation}
for all invariants $s_{ij}=(p_{i}+p_{j})^{2}$, where $p_{i}$ and
$p_{j}$ are the 4-momenta of partons $i$ and $j$, respectively.
For the massive quarks, we follow the definition in Ref.~\citep{Brandenburg:1997pu}
to account for masses, but still use the terminology {}``resolved''
and {}``unresolved'' partons. In the regions with unresolved partons,
soft and collinear approximations of the matrix elements, which hold
exactly in the limit $s_{min}\rightarrow0$, are used. The necessary
integrations over the soft and collinear regions of phase space can
then be carried out analytically in $d=4-2\epsilon$ space-time dimensions.
One can thus isolate all the poles in $\epsilon$ and perform the
cancellation of the IR singularities between the real and virtual
contributions and absorb the leftover singularities into the parton
structure functions in the factorization procedure. After the above
procedure, one takes the limit $\epsilon\rightarrow0$. The contribution
from the sum of virtual and unresolved region corrections is finite
but $s_{min}$ dependent. Since the parameter $s_{min}$ is introduced
in the theoretical calculation for technical reasons only and is unrelated
to any physical quantity, the sum of all contributions (virtual, unresolved
and resolved corrections) must not depend on $s_{min}$. The phase
space slicing method is only valid in the limit that $s_{{\rm min}}$
is small enough that a given jet finding algorithm (or any infrared-safe
observable) can be consistently defined even after including the experimental
cuts.

\begin{figure}
\includegraphics[clip,scale=0.68]{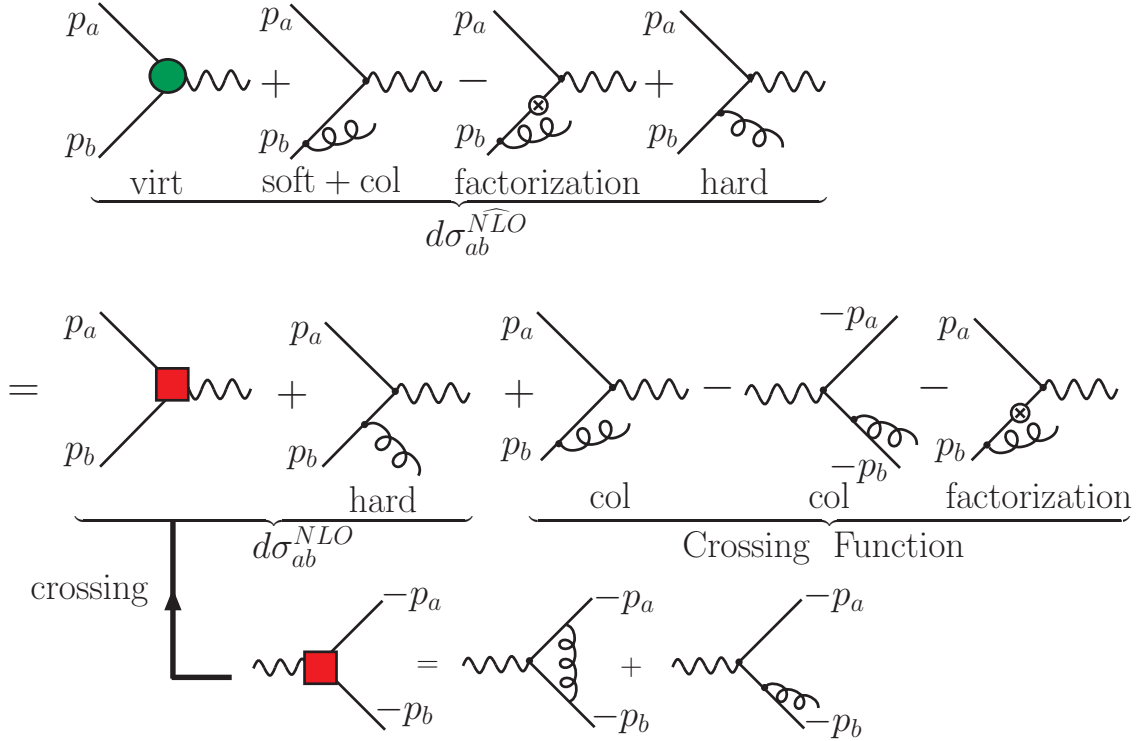}

\caption{Illustration of the PSS method with one cutoff scale to describe the
processes with initial state massless quarks. Here, only half of the
real emission diagrams is shown. In this paper, we will assign the
particle's momentum such that the initial state particle's momentum
is incoming to the vertex while the final state particle's momentum
is outgoing.\label{fig:crossing}}

\end{figure}

In general, the conventional calculation of the NLO differential cross
section for a process with initial state hadrons $H_{1}$ and $H_{2}$
can be written as\begin{equation}
d\sigma_{H_{1}H_{2}}^{NLO}=\sum_{a,b}\int dx_{1}dx_{2}f_{a}^{H_{1}}(x_{1},\mu_{F})f_{b}^{H_{2}}(x_{2},\mu_{F})d\widehat{\sigma_{ab}^{NLO}}(x_{1,}x_{2},\mu_{R}),\label{eq:conventional_NLO}\end{equation}
where $a$, $b$ denote parton flavors and $x_{1}$,$x_{2}$ are parton
momentum fractions. $f_{a}^{H}(x,\mu_{F})$ is the usual NLO parton
distribution function with the mass factorization scale $\mu_{F}$
and $d\widehat{\sigma_{ab}^{NLO}}(x_{1},x_{2},\mu_{R})$ is the NLO
hard scattering differential cross section with the renormalization
scale $\mu_{R}$. The pictorial demonstration of Eq.~(\ref{eq:conventional_NLO})
is shown in the upper part of Fig.~\ref{fig:crossing}. 

Contrary to the conventional calculation method, the PSS method with
one cutoff scale will firstly cross the initial state partons into
the final state, including the virtual corrections and unresolved
real emission corrections. For example, to calculate the NLO QCD correction
to the $W$-boson production via the Drell-Yan process, we first calculate
the radiative corrections to $W\rightarrow q\bar{q}'(g)$, as shown
in the lower part of the Fig.~\ref{fig:crossing}, in which we split
the phase space of the real emission corrections into the unresolved
and resolved region. After we integrate out the unresolved phase space
region, the net contribution of the virtual corrections and the real
emission corrections in the unresolved phase space is finite but theoretical
cutoff $s_{min}$ dependent, which can be written as a form factor
(denoted by the box in Fig.~\ref{fig:crossing}) of the Born level
vertex.

Secondly, we take the already calculated effective matrix elements
with all the partons in the final state and use the universal {}``crossing
function'', which is the generalization of the crossing property
of the LO matrix elements to NLO, to calculate the corresponding matrix
elements numerically. Once we cross the needed partons to the initial
state, the contributions from the unresolved collinear phase space
regions are different from those with all the partons in the final
state. These differences are included into the definition of the crossing
function as well as the mass factorization effects, as shown in the
middle part of Fig.~\ref{fig:crossing}. Here, we only present the
explicit expressions of the crossing function, while the definition
and detailed derivation of the crossing function can be found in Ref.~\citep{Giele:1993dj}.
After applying the mass factorization in a particular scheme, the
crossing functions for an initial state parton $a$, which participates
in the hard scattering processes, can be written in the form:\begin{equation}
C_{a}^{{\rm scheme}}(x,\mu_{F},s_{min})=\left(\frac{N_{C}}{2\pi}\right)\left[A_{a}(x,\mu_{F})\log\left(\frac{s_{min}}{\mu_{F}}\right)+B_{a}^{{\rm scheme}}(x,\mu_{F})\right],\label{eq:crossing_def}\end{equation}
where\begin{eqnarray}
A_{a}(x,\mu_{F}) & = & \sum_{p}A_{p\rightarrow a}\left(x,\mu_{F}\right),\label{eq:crossing_def_a}\\
B_{a}^{{\rm scheme}}(x,\mu_{F}) & = & \sum_{p}B_{p\rightarrow a}^{{\rm scheme}}(x,\mu_{F}),\label{eq:crossing_def_b}\end{eqnarray}
and $N_{C}$ denotes the number of colors. The sum runs over $p=q,\bar{q},g$.
The functions $A$ and $B$ can be expressed as convolution integrals
over the parton distribution functions and the explicit forms can
be found in Ref.~\citep{Giele:1993dj}. Although $A_{a}$ is scheme
independent, $B_{a}$ does depend on the mass factorization scheme,
and therefore so does the crossing function. 

After introducing the crossing function, we can write the NLO differential
cross section in the PSS method with one cutoff scale as\begin{eqnarray}
 &  & d\sigma_{H_{1}H_{2}}^{NLO}\nonumber \\
 & = & \sum_{a,b}\int dx_{1}dx_{2}f_{a}^{H_{1}}(x_{1},\mu_{F})f_{b}^{H_{2}}(x_{2},\mu_{F})d\sigma_{ab}^{NLO}(x_{1},x_{2},\mu_{R})\nonumber \\
 & + & \alpha_{s}(\mu_{R})\left[C_{a}^{H_{1}}(x_{1},\mu_{F})f_{b}^{H_{2}}(x_{2},\mu_{F})+f_{a}^{H_{1}}(x_{1},\mu_{F})C_{b}^{H_{2}}(x_{2},\mu_{F})\right]d\sigma_{ab}^{LO}(x_{1},x_{2}).\label{eq:nlo-formalism-1}\end{eqnarray}
Here $d\sigma_{ab}^{NLO}$ consists of the finite effective all-partons-in-the-final-state
matrix elements, in which partons $a$ and $b$ have simply been crossed
to the initial state, i.e. in which their momenta $-p_{a}$ and -$p_{b}$
have been replaced by $p_{a}$ and $p_{b}$, as shown in the Fig.~\ref{fig:crossing}.
The difference between $d\sigma_{ab}^{NLO}$ and $d\widehat{\sigma_{ab}^{NLO}}$
has been absorbed into the finite, universal crossing function $C_{a}^{H}(x,\mu_{F})$.
Defining a {}``effective'' NLO parton distribution function $\mathcal{F}_{a}^{H}(x)$
as\begin{equation}
\mathcal{F}_{a}^{H}(x)=f_{a}^{H}(x,\mu_{F})+\alpha_{s}(\mu_{R})C_{a}^{H}(x,\mu_{F})+O(\alpha_{s}^{2}),\label{eq:effective-pdf}\end{equation}
we can rewrite Eq.~(\ref{eq:nlo-formalism-1}) in a simple form as\begin{equation}
d\sigma_{H_{1}H_{2}}^{NLO}=\sum_{a,b}\int dx_{1}dx_{2}\mathcal{F}_{a}^{H_{1}}(x_{1})\mathcal{F}_{b}^{H_{2}}(x_{2})d\sigma_{ab}^{NLO}(x_{1},x_{2}).\label{eq:PSS-NLO-Formalisam}\end{equation}

\section{Tree level matrix elements}

\begin{figure}
\includegraphics[clip,scale=0.6]{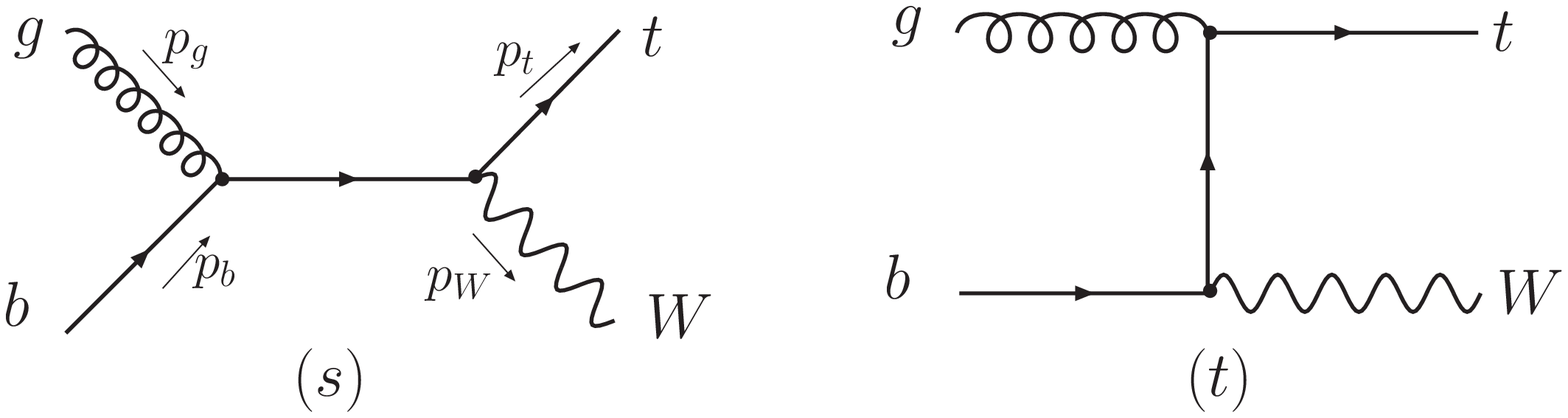}

\caption{Tree-level Feynman diagram of the $gb\to tW$ process.\label{fig:twtree}}

\end{figure}

The matrix elements of the scattering process $gb\to tW$ can be written
in terms of the following twenty standard matrix elements that contain
the information about the Dirac matrix structure:\begin{equation}
\mathcal{M}_{LO}=\mathcal{M}_{LO}^{(s)}+\mathcal{M}_{LO}^{(t)},\end{equation}
where \begin{eqnarray}
 &  & \mathcal{M}_{LO}^{(i)}=-i\frac{g_{s}g}{\sqrt{2}}T_{ij}^{a}\epsilon^{\mu}(g)\varepsilon^{\nu*}(W)\,\bar{u}_{t}\biggl[f_{1}^{i}g_{\mu\nu}+f_{2}^{i}\gamma_{\mu}\gamma_{\nu}+f_{3}^{i}p_{b,\nu}\gamma_{\mu}+f_{4}^{i}p_{b,\mu}\gamma_{\nu}+f_{5}^{i}p_{t,\nu}\gamma_{\mu}\nonumber \\
 &  & \qquad\qquad\qquad+f_{6}^{i}p_{t,\mu}\gamma_{\nu}+f_{7}^{i}p_{b,\mu}p_{b,\nu}+f_{8}^{i}p_{b,\mu}p_{t,\nu}+f_{9}^{i}p_{t,\mu}p_{b,\nu}+f_{10}^{i}p_{t,\mu}p_{t,\nu}\nonumber \\
 &  & \qquad+\pslash{p}_{W}(f_{11}^{i}g_{\mu\nu}+f_{12}^{i}\gamma_{\mu}\gamma_{\nu}+f_{13}^{i}p_{b,\nu}\gamma_{\mu}+f_{14}^{i}p_{b,\mu}\gamma_{\nu}+f_{15}^{i}p_{t,\nu}\gamma_{\mu}\nonumber \\
 &  & \qquad\qquad\qquad+f_{16}^{i}p_{t,\mu}\gamma_{\nu}+f_{17}^{i}p_{b,\mu}p_{b,\nu}+f_{18}^{i}p_{b,\mu}p_{t,\nu}+f_{19}^{i}p_{t,\mu}p_{b,\nu}+f_{20}^{i}p_{t,\mu}p_{t,\nu})\biggr]P_{L}u_{b},\label{eq:formfactor}\end{eqnarray}
where $i=s(t)$. Here, $g_{s}(g)$ denotes the coupling strength of
the strong (weak) interaction, respectively, $T_{i,j}^{a}$ are the
color matrices in the fundamental representation, and $p_{L}\equiv(1-\gamma_{5})/2$
denotes the left-handed projector. The tree-level Feynman diagrams
of the $gb\to tW$ process are shown in Fig.\ \ref{fig:twtree}.
The non-zero form factors of the tree-level matrix elements are given
by\begin{eqnarray}
 &  & -f_{3}^{s}=f_{4}^{s}=f_{5}^{s}=f_{11}^{s}=2f_{12}^{s}=\frac{2}{s},\\
 &  & f_{6}^{t}=2f_{12}^{t}=\frac{2}{t_{1}}.\end{eqnarray}
The kinematics variables used in this paper are defined as follows:\begin{eqnarray}
 &  & s=\left(p_{g}+p_{b}\right)^{2}=2p_{g}\cdot p_{b},\\
 &  & t=\left(p_{g}-p_{t}\right)^{2}=t_{1}+m_{t}^{2},\\
 &  & u=\left(p_{g}-p_{W}\right)^{2}=u_{1}+m_{W}^{2},\end{eqnarray}
where $t_{1}\equiv-2p_{g}\cdot p_{t}$, $u_{1}\equiv-2p_{g}\cdot p_{W}$,
and $s+t_{1}+u_{1}=0$.

\section{Virtual corrections and renormalization}

\begin{figure}
\includegraphics[clip,scale=0.7]{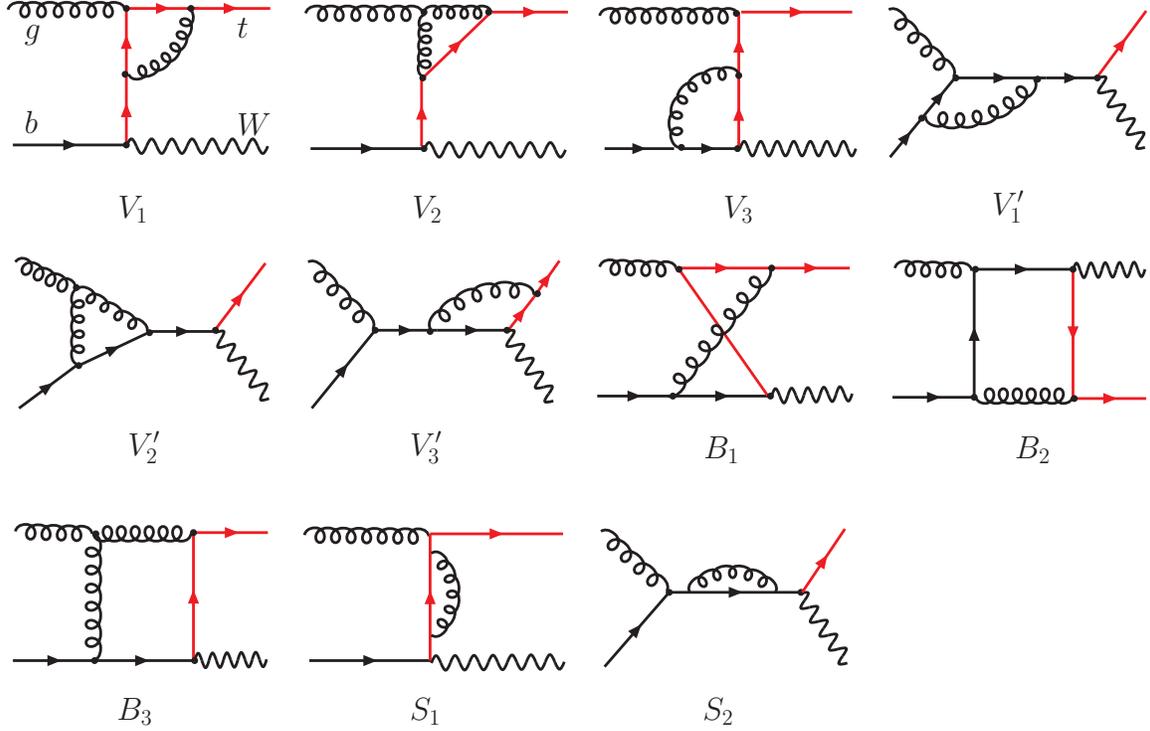}

\caption{Feynman diagrams of the one-loop virtual corrections where the red
lines denote the top quark.\ \label{fig:twvirtual}}

\end{figure}

Now let us calculate the virtual corrections to the $tW$ associated
production. Fig.\ \ref{fig:twvirtual} shows the Feynman diagrams
of the one-loop QCD virtual corrections, where $V_{i}$ denote the
triangle loop corrections, $B_{i}$ the box loop corrections, while
$S_{i}$ the bubble loop corrections. In this work we adapt the on-shell
renormalization scheme such that the external self-energy corrections
are cancelled by a set of counterterms which will be discussed later
in this sections. 

\begin{figure}
\includegraphics[clip,scale=0.6]{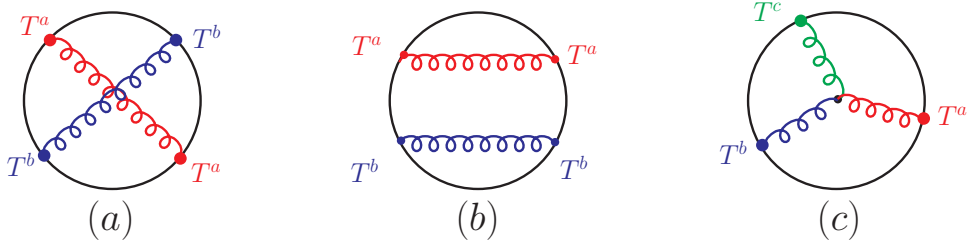}

\caption{Color factor of the interference of tree-level and one-loop matrix
element.\label{fig:twvitual_colorfactor}}

\end{figure}

At the NLO the relevant one-loop virtual corrections contribute only
through their interference with the lowest-order Born amplitudes.
The interference, i.e. $\Re\left(\mathcal{M}_{0}\mathcal{M}_{Virtual}^{\dagger}\right)$,
gives rise to the order of $\alpha_{s}^{2}$ contributions. One can
further categorize the virtual corrections into three classes, according
to their color structures depicted in Fig.\ \ref{fig:twvitual_colorfactor}:
(a) $V_{1},\, V_{1}^{\prime},\, B_{1},\, B_{2}$; (b) $V_{3},\, V_{3}^{\prime},\, S_{1},\, S_{2}$;
(c) $V_{2},\, V_{2}^{\prime},\, B_{3}$. The color factors (CF) of
those three classes, normalized to the Born level matrix element square,
are given by\begin{equation}
CF_{a}=-\frac{1}{2N_{C}},\quad CF_{b}=\frac{N_{C}}{2}-\frac{1}{2N_{C}},\quad CF_{c}=\frac{N_{C}}{2}.\end{equation}
The same color structures also apply to the real radiation corrections.
Below we will make use of the color order to organize our calculation. 

To calculate the virtual corrections, we follow the Passarino-Veltman
procedure\ \citep{Passarino:1978jh,'tHooft:1978xw}: one can reduce
any set of vectorial of tensorial one-loop integrals to a few scalar
functions. However, since the loop corrections exhibit not only the
UV divergence but also the IR divergence, either soft or collinear
or both, one needs to work out those scalar function analytically.
Recently, a basis set of infra-red and/or collinearly divergent scalar
oneloop integrals is constructed, and analytic formulas, for tadpole,
bubble, triangle and box integrals, are given explicitly in Ref.\ \citep{Ellis:2007qk}.
The scalar functions used in this work can be found in that paper
and those results are confirmed by our hand calculation. Our final
results of the virtual corrections are very complicated, but the divergent
pieces can be factorized out in a simple form. We present the full
expressions and the divergent poles of the form factors in the Appendix\ \ref{sec:formfactor}
and \ref{sec:Divergences-of-formfactor}, respectively. We further
distinguish between the UV and IR divergences in order to keep track
on the renormalization and factorization. 

Consider now the UV divergence first. We define $\Delta_{UV}(\mathcal{M}_{i})$
to be the UV pole part of the corresponding amplitude $\mathcal{M}_{i}$.
Using this notation, we find\begin{eqnarray}
\Delta_{UV}\left(\mathcal{M}_{V_{1}}+\mathcal{M}_{V_{1}^{\prime}}\right) & = & \frac{\alpha_{s}}{4\pi}C_{\epsilon}\left(-\frac{1}{2N_{C}}\right)\frac{1}{\epsilon_{UV}}\mathcal{M}_{LO},\\
\Delta_{UV}\left(\mathcal{M}_{V_{2}}+\mathcal{M}_{V_{2}^{\prime}}\right) & = & \frac{\alpha_{s}}{4\pi}C_{\epsilon}\left(\frac{N_{C}}{2}\right)\frac{3}{\epsilon_{UV}}\mathcal{M}_{LO},\\
\Delta_{UV}\left(\mathcal{M}_{V_{3}}+\mathcal{M}_{V_{3}^{\prime}}\right) & = & \frac{\alpha_{s}}{4\pi}C_{\epsilon}\left(\frac{N_{C}}{2}-\frac{1}{2N_{C}}\right)\frac{1}{\epsilon_{UV}}\mathcal{M}_{LO},\\
\Delta_{UV}\left(\mathcal{M}_{S_{1}}+\mathcal{M}_{S_{2}}\right) & = & \frac{\alpha_{s}}{4\pi}C_{\epsilon}\left(\frac{N_{C}}{2}-\frac{1}{2N_{C}}\right)\frac{-1}{\epsilon_{UV}}\mathcal{M}_{LO}+\mathcal{A}_{S_{1}},\label{eq:div-s}\end{eqnarray}
where $C_{\epsilon}\equiv(4\pi\mu^{2}/m_{t}^{2})^{2}\Gamma\left(1+\epsilon\right)$.
Note that $\mathcal{A}_{S_{1}}$ in Eq.\ \ref{eq:div-s} represents
the non-factorization loop corrections originated from the Feynman
diagram $S_{1}$, see Eqs.\ (\ref{eq:fs1_2}-\ref{eq:fs1_12}). The
UV divergent poles of $\mathcal{A}_{S_{1}}$ are given in terms of
the form factors as follows:\begin{equation}
f_{2}^{\mathcal{A}_{S1}}=\frac{m_{t}}{t_{1}}\frac{3}{\epsilon_{UV}},\qquad f_{6}^{\mathcal{A}_{S1}}=2f_{12}^{\mathcal{A}_{S1}}=\frac{1}{\epsilon_{UV}}\left(\frac{12m_{t}^{2}}{t_{1}^{2}}-\frac{2}{t_{1}}\right).\end{equation}
As to be shown later, these non-factorisable divergences will be exactly
cancelled by the top quark mass renormalization. 

The renormalization is preformed in the $\overline{{\rm MS}}$ scheme
with the top quark mass defined on shell. As required by renormalization
group arguments, the renormalization of the fundamental propagators
and interaction vertices of the theory reduces to introducing counterterms
for the external field wave functions of top quarks and gluons ($\delta Z_{t}$,
$\delta Z_{G}$), for the top mass ($\delta m_{t}$), and for the
strong coupling constant ($\delta g_{s}$). We renormalize the fields
of the gluons, $G_{a,\mu}$, of the bottom quark, $\psi_{b}$, and
of the top quark, $\psi_{t}$, all in the on-shell scheme, i.e. the
wave-function renormalization constants $\delta Z_{G,b,t}$, defined
by the transformations\begin{equation}
G_{a,\mu,}^{0}=\left(1+\frac{1}{2}\delta Z_{G}\right)G_{a,\mu},\quad\psi_{b}^{0}=\left(1+\frac{1}{2}\delta Z_{b}\right)\psi_{b},\quad\psi_{t}^{0}=\left(1+\frac{1}{2}Z_{t}\right)\psi_{t},\end{equation}
are adjusted to cancel the external self-energy corrections exactly.
Distinguishing between divergences of UV and IR origin, these constants
can be written as\begin{eqnarray}
\delta Z_{G} & = & -\frac{\alpha_{s}}{4\pi}C_{\epsilon}\left\{ \left(\frac{2}{3}n_{f}-\frac{5N_{C}}{3}\right)\left(\frac{1}{\epsilon_{UV}}-\frac{1}{\epsilon_{IR}}\right)-\frac{2}{3}\frac{1}{\epsilon_{UV}}\right\} ,\\
\delta Z_{t} & = & -\frac{\alpha_{s}}{4\pi}C_{\epsilon}\left(\frac{N_{C}}{2}-\frac{1}{2N_{C}}\right)\left(\frac{1}{\epsilon_{UV}}+\frac{2}{\epsilon_{IR}}+4\right),\\
\delta Z_{b} & = & -\frac{\alpha_{s}}{4\pi}C_{\epsilon}\left(\frac{N_{C}}{2}-\frac{1}{2N_{C}}\right)\left(\frac{1}{\epsilon_{UV}}-\frac{1}{\epsilon_{IR}}\right),\end{eqnarray}
where $n_{f}=5$ is the number of light quark flavours. 

Denoting the bare top quark mass and the bare strong coupling as $m_{t}^{0}$
and $\alpha_{s}^{0}$, respectively, we introduce the renormalization
parameters in the transformations\begin{equation}
m_{t}^{0}=m_{t}+\delta m_{t},\qquad\alpha_{s}^{0}=\alpha_{s}+\delta\alpha_{s}.\end{equation}
We defined the subtraction condition for the top quark mass $m_{t}$
in such a way that $m_{t}$ is the pole mass, in which case the top
mass counterterm is given by\begin{equation}
\frac{\delta m_{t}}{m_{t}}=-\frac{\alpha_{s}}{4\pi}C_{\epsilon}\left(\frac{N_{C}}{2}-\frac{1}{2N_{C}}\right)\left(\frac{3}{\epsilon_{UV}}+4\right).\end{equation}
Finally, for the renormalization of $\alpha_{s}$, we use the $\overline{{\rm MS}}$
scheme, modified to decouple the top quark from the running of the
strong coupling $\alpha_{s}\left(\mu\right)$. It gives rise to\[
\frac{\delta\alpha_{s}}{\alpha_{s}}=\frac{\alpha_{s}}{4\pi}\left\{ \left(4\pi\right)^{\epsilon}\Gamma\left(1+\epsilon\right)\left(\frac{2n_{f}}{3}-\frac{11N_{C}}{3}\right)\frac{1}{\epsilon_{UV}}+\frac{2}{3}C_{\epsilon}\frac{1}{\epsilon_{UV}}\right\} ,\]
where the first term inside the braces originates from light quark
and gluon loops, which is independent of $\mu$, while the second
term (proportional to $C_{\epsilon}$) originates from the top quark
loop in the gluon self-energy that is subtracted at zero-momentum
transfer. In such a way the running of the coupling $\alpha_{s}\left(\mu\right)$
is generated solely by the finite contributions of the light quark
and gluon loops, while the top quark contribution is absorbed completely
in the renormalization condition and thus decouple effectively. 

\begin{figure}
\includegraphics[clip,scale=0.6]{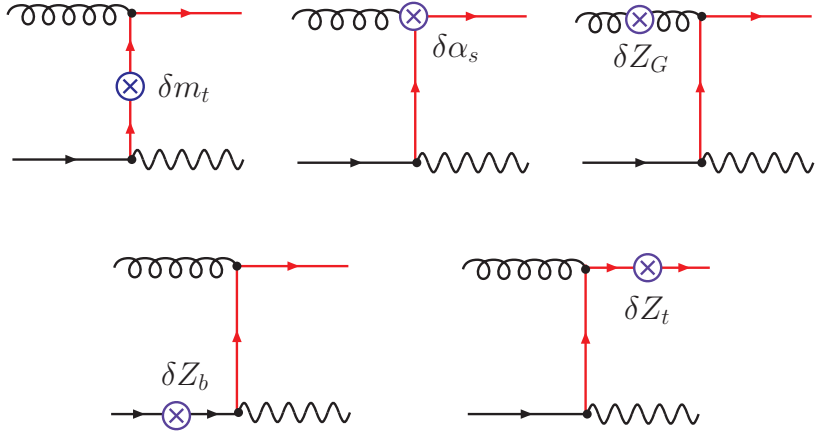}

\caption{Depicted Feynman diagrams of the counterterm contributions.\label{fig:conterterm}}

\end{figure}

The contributions of the counterterms and the external self-energy
to the one-loop matrix element are depicted in Fig.\ \ref{fig:conterterm}.
The matrix elements is given by\begin{equation}
\left(\frac{\delta\alpha_{s}}{\alpha_{s}}+\frac{1}{2}\delta Z_{b}+\frac{1}{2}\delta Z_{t}+\frac{1}{2}\delta Z_{G}\right)\mathcal{M}_{LO}+\mathcal{M}(\delta m_{t}).\end{equation}
The first term accounts for the external self-energy corrections as
well as the counterterms of the strong coupling renormalization, while
the second term proportional to $\delta m_{t}$ comes from the renormalization
of the top quark mass. After insertion of the top quark mass counterterm,
see Fig.\ \ref{fig:conterterm}, we obtain the form factor of $\mathcal{M}\left(\delta m_{t}\right)$
as follows (where the color factor $\frac{N}{2}-\frac{1}{2N}$ is
not shown explicitly):\begin{eqnarray}
f_{2}^{\delta m_{t}} & = & -\frac{m_{t}}{t_{1}}\left[\frac{3}{\epsilon_{UV}}+4\right],\\
f_{6}^{\delta m_{t}} & = & -\frac{m_{t}^{2}}{t_{1}^{2}}\left[\frac{12}{\epsilon_{UV}}+16\right],\\
f_{12}^{\delta m_{t}} & = & \frac{1}{2}f_{6}^{\delta m_{t}}.\end{eqnarray}

In order to check the renormalization, we will combine $\mathcal{M}\left(\delta m_{t}\right)$
with those bubble, triangle and box loop corrections, therefore the
renormalized matrix elements are given by\begin{eqnarray}
\mathcal{M}_{ren} & = & \mathcal{M}_{virt}+\mathcal{M}_{CT},\\
\mathcal{M}_{virt} & = & \left[\sum_{i=1}^{2}\mathcal{M}_{S_{i}}+\sum_{i=1}^{3}\left(\mathcal{M}_{V_{i}}+\mathcal{M}_{V_{i}^{\prime}}+\mathcal{M}_{B_{i}}\right)+\mathcal{M}\left(\delta m_{t}\right)\right],\\
\mathcal{M}_{CT} & = & \left(\frac{\delta\alpha_{s}}{\alpha_{s}}+\frac{1}{2}\delta Z_{b}+\frac{1}{2}\delta Z_{t}+\frac{1}{2}\delta Z_{G}\right)\mathcal{M}_{LO}.\end{eqnarray}
It is straightforward to check the UV divergences indeed cancel out,
but the IR (soft and collinear) divergences still remain to be cancelled
by the real radiation corrections. The IR poles of the virtual corrections
are given by\begin{eqnarray}
\Delta_{IR}^{virt} & = & \mathcal{M}_{LO}\times\frac{\alpha_{s}}{4\pi}C_{\epsilon}\biggl\{-\frac{13}{\epsilon_{IR}^{2}}+\frac{3}{\epsilon_{IR}}\ln\frac{-t_{1}}{m_{t}^{2}}+\frac{3}{\epsilon_{IR}}\ln\frac{s}{m_{t}^{2}}-\frac{1}{3\epsilon_{IR}}\ln\frac{m_{t}^{2}-m_{W}^{2}-u}{m_{t}^{2}}\nonumber \\
 &  & \qquad\qquad\qquad\,\,\,\,-\frac{10}{3\epsilon_{IR}}-\frac{1}{2\epsilon_{IR}}\left(\frac{11N_{C}}{3}-\frac{2}{3}n_{f}\right)\biggr\}.\end{eqnarray}

\section{Infra-red singularities in the real emission corrections}

In the one cutoff phase space slicing method, the theoretical cutoff
parameter ($s_{min}$) is introduced in order to isolate soft and
collinear singularities associated with real gluon emission sub-processes
by partitioning the phase space into soft, collinear and hard regions
such that\begin{equation}
\left|\mathcal{M}^{{\rm r}}\right|^{2}=\left|\mathcal{M}^{{\rm r}}\right|_{{\rm soft}}^{2}+\left|\mathcal{M}^{{\rm r}}\right|_{{\rm collinear}}^{2}+\left|\mathcal{M}^{{\rm r}}\right|_{{\rm hard}}^{2}\,.\label{eq:nlomat}\end{equation}
In the soft and collinear regions the cross section is proportional
to the Born-level cross section. Using dimensional regularization,
we can evaluate the real gluon emission diagrams in n-dimensions under
the soft gluon approximation in the soft region, or the collinear
approximation in the collinear region, and can integrate out the corresponding
phase space volume analytically. The resulting divergences are cancelled
by virtual corrections or absorbed into the perturbative parton distribution
functions in the factorization procedure. 

In this method, a pair of partons with momenta $p_{i}$ and $p_{j}$
is defined to be unresolved if\begin{equation}
\left|2p_{i}\cdot p_{j}\right|<s_{min},\label{eq:condition2}\end{equation}
with $s_{min}$ small compared to the hard scale of the process. This
condition can occur if either $p_{i}$ and $p_{j}$ are collinear,
or if one of the two is soft. When the scattering amplitude involves
a complicated color structure, one needs to decompose the scattering
amplitude into color-ordered sub-amplitudes. Below we will use the
$tW$ associated production process to illustrate this point. 

\begin{figure}
\includegraphics[scale=0.4]{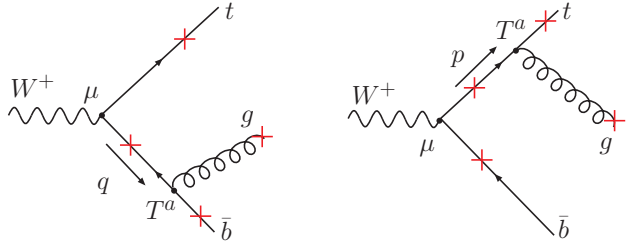}

\caption{Feynman diagrams of process $W'\to t\bar{b}g$ at Born level.\label{fig:born_w2qqg}}

\end{figure}

To calculate the real emission corrections, we first cross all the
initial state partons to the final state and the final state $W$
boson into the initial state, i.e. at the tree level,\begin{eqnarray*}
gb\to W^{-}t & \xrightarrow{crossing} & W^{+}\to t\bar{b}g.\end{eqnarray*}
The tree-level Feynman diagrams after crossing are shown in Fig.~\ref{fig:born_w2qqg},
which give rise to the following matrix element\begin{equation}
\mathcal{M}_{0}=-i\frac{g}{\sqrt{2}}g_{s}T^{a}\epsilon^{\mu}(W)\epsilon^{\sigma*}(g)\bar{u}(t)\left[\gamma^{\sigma}\frac{\pslash{p}+m_{t}}{p^{2}-m_{t}^{2}}\gamma^{\mu}P_{L}+\gamma^{\mu}P_{L}\frac{-\pslash{q}+m_{b}}{q^{2}-m_{b}^{2}}\gamma^{\sigma}\right]v(\bar{b}),\end{equation}
where $q=p_{g}+p_{\bar{b}}$ and $p=p_{t}+p_{g}$. The cross symbol
{}``$\times$'' in the figure indicates the possible places where
the additional parton can be radiated from. There are two types of
real radiation corrections:\[
(1)\, W^{+}\to t\bar{b}gg;\qquad(2)\, W^{+}\to t\bar{b}q\bar{q}.\]
The former exhibits both soft and collinear divergences, but the latter
can only have the collinear divergence.

\subsection{Color ordered amplitude}

\begin{figure}
\includegraphics[scale=0.7]{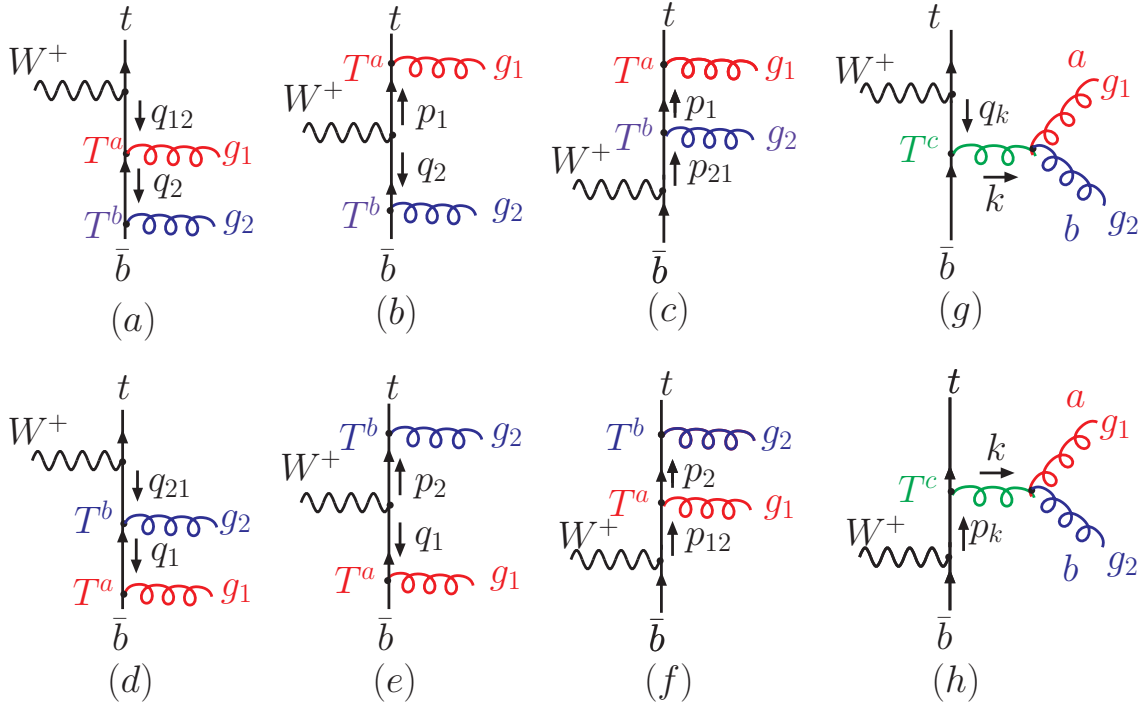}

\caption{Feynman diagrams contributing to $t\bar{b}gg$ final state. \label{fig:color_order}}

\end{figure}

For a systematic extraction of the infrared singularities within the
one cutoff method, we organize the amplitude in terms of color-ordered
sub-amplitudes. In Fig.~\ref{fig:color_order}, we present all the
real emission diagrams which give rise to the final state of $t\bar{b}gg$.
The color coefficients of the diagrams are given as follows:\begin{eqnarray*}
a,b,c & \propto & T^{a}T^{b},\\
d,e,f & \propto & T^{b}T^{a},\\
g,h\,\, & \propto & T^{a}T^{b}\,\,{\rm and}\,\, T^{b}T^{a},\end{eqnarray*}
where we have applied the identity\begin{equation}
if^{abc}T^{c}=T^{a}T^{b}-T^{b}T^{c},\end{equation}
in the diagrams (g) and (h). Thus, we can decompose the amplitude
$\mathcal{M}(W^{+}\to t\bar{b}gg)$ as following,\begin{equation}
\mathcal{M}(W^{+}\to t\bar{b}gg)=T^{a}T^{b}\mathcal{M}_{1}+T^{b}T^{a}\mathcal{M}_{2},\label{eq:color_ordered_amp}\end{equation}
where\begin{eqnarray*}
\mathcal{M}_{1} & = & \mathcal{M}^{(a)}+\mathcal{M}^{(b)}+\mathcal{M}^{(c)}-\mathcal{M}^{(g)}-\mathcal{M}^{(h)},\\
\mathcal{M}_{2} & = & \mathcal{M}^{(d)}+\mathcal{M}^{(e)}+\mathcal{M}^{(f)}+\mathcal{M}^{(g)}+\mathcal{M}^{(h)}.\end{eqnarray*}
The color indices are assigned as shown in Fig.\ \ref{fig:color_order}.
The momentum $q_{i}$, $q_{ij}$, $p_{i}$ and $p_{ij}$ are defined
as\begin{eqnarray*}
 &  & p_{i}=p_{t}+p_{g_{i}},\quad p_{ji}=p_{t}+p_{g_{i}}+p_{g_{j}},\\
 &  & q_{i}=p_{b}+p_{g_{i}},\quad q_{ji}=p_{b}+p_{g_{i}}+p_{g_{j}}.\end{eqnarray*}

The amplitudes of diagrams (a, b, c) are given as follows (with a
color coefficient $T^{a}T^{b}$):\begin{eqnarray}
\mathcal{M}^{(a)} & = & i\frac{g}{\sqrt{2}}g_{s}^{2}\epsilon^{\mu}(W^{+})\epsilon^{\rho*}(g_{1})\epsilon^{\sigma*}(g_{2})\nonumber \\
 & \times & \bar{u}(t)\left[\gamma_{\mu}P_{L}\frac{-\pslash{q_{12}}+m_{b}}{q_{12}^{2}-m_{b}^{2}}\gamma_{\rho}\frac{-\pslash{q_{2}}+m_{b}}{q_{2}^{2}-m_{b}^{2}}\gamma_{\sigma}\right]v\left(\bar{b}\right),\\
\mathcal{M}^{(b)} & = & i\frac{g}{\sqrt{2}}g_{s}^{2}\epsilon^{\mu}(W^{+})\epsilon^{\rho*}(g_{1})\epsilon^{\sigma*}(g_{2})\nonumber \\
 & \times & \bar{u}(t)\left[\gamma_{\rho}\frac{\pslash{p_{1}}+m_{t}}{p_{1}^{2}-m_{t}^{2}}\gamma_{\mu}P_{L}\frac{-\pslash{q_{2}}+m_{b}}{q_{2}^{2}-m_{b}^{2}}\gamma_{\sigma}\right]v\left(\bar{b}\right),\\
\mathcal{M}^{(c)} & = & i\frac{g}{\sqrt{2}}g_{s}^{2}\epsilon^{\mu}(W^{+})\epsilon^{\rho*}(g_{1})\epsilon^{\sigma*}(g_{2})\nonumber \\
 & \times & \bar{u}(t)\left[\gamma_{\rho}\frac{\pslash{p_{1}}+m_{t}}{p_{1}^{2}-m_{t}^{2}}\gamma_{\sigma}\frac{\pslash{p_{21}}+m_{t}}{p_{21}^{2}-m_{t}^{2}}\gamma_{\mu}P_{L}\right]v\left(\bar{b}\right).\end{eqnarray}
The amplitudes of diagrams (d, e, f) are given as follows (with a
color coefficient $T^{b}T^{a}$):\begin{eqnarray}
\mathcal{M}^{(d)} & = & i\frac{g}{\sqrt{2}}g_{s}^{2}\epsilon^{\mu}(W^{+})\epsilon^{\rho*}(g_{1})\epsilon^{\sigma*}(g_{2})\nonumber \\
 & \times & \bar{u}(t)\left[\gamma_{\mu}P_{L}\frac{-\pslash{q_{21}}+m_{b}}{q_{21}^{2}-m_{b}^{2}}\gamma_{\sigma}\frac{-\pslash{q_{1}}+m_{b}}{q_{1}^{2}-m_{b}^{2}}\gamma_{\rho}\right]v\left(\bar{b}\right),\\
\mathcal{M}^{(e)} & = & i\frac{g}{\sqrt{2}}g_{s}^{2}\epsilon^{\mu}(W^{+})\epsilon^{\rho*}(g_{1})\epsilon^{\sigma*}(g_{2})\nonumber \\
 & \times & \bar{u}(t)\left[\gamma_{\sigma}\frac{\pslash{p_{2}}+m_{t}}{p_{2}^{2}-m_{t}^{2}}\gamma_{\mu}P_{L}\frac{-\pslash{q_{1}}+m_{b}}{q_{1}^{2}-m_{b}^{2}}\gamma_{\rho}\right]v\left(\bar{b}\right),\\
\mathcal{M}^{(f)} & = & i\frac{g}{\sqrt{2}}g_{s}^{2}\epsilon^{\mu}(W^{+})\epsilon^{\rho*}(g_{1})\epsilon^{\sigma*}(g_{2})\nonumber \\
 & \times & \bar{u}(t)\left[\gamma_{\sigma}\frac{\pslash{p_{2}}+m_{t}}{p_{2}^{2}-m_{t}^{2}}\gamma_{\rho}\frac{\pslash{p_{12}}+m_{t}}{p_{12}^{2}-m_{t}^{2}}\gamma_{\mu}P_{L}\right]v\left(\bar{b}\right).\end{eqnarray}
Finally, the amplitudes of diagrams (g, h) are given as follows (with
a color coefficient $-if^{abc}T^{c}$):\begin{eqnarray}
\mathcal{M}^{(g)} & = & i\frac{g}{\sqrt{2}}g_{s}^{2}\epsilon^{\mu}(W^{+})\epsilon^{\rho*}(g_{1})\epsilon^{\sigma*}(g_{2})\nonumber \\
 & \times & \bar{u}(t)\left[\gamma_{\mu}P_{L}\frac{-\pslash{q_{k}}+m_{b}}{q_{k}^{2}-m_{b}^{2}}\gamma_{\alpha}\right]v\left(\bar{b}\right)\frac{1}{k^{2}}V^{\rho\sigma\alpha},\\
\mathcal{M}^{(h)} & = & i\frac{g}{\sqrt{2}}g_{s}^{2}\epsilon^{\mu}(W^{+})\epsilon^{\rho*}(g_{1})\epsilon^{\sigma*}(g_{2})\nonumber \\
 & \times & \bar{u}(t)\left[\gamma_{\alpha}\frac{\pslash{p_{k}}+m_{t}}{p_{k}^{2}-m_{t}^{2}}\gamma_{\mu}P_{L}\right]v\left(\bar{b}\right)\frac{1}{k^{2}}V^{\rho\sigma\alpha},\end{eqnarray}
where\begin{equation}
V^{\rho\sigma\alpha}=g^{\rho\sigma}\left(p_{g_{1}}-p_{g_{2}}\right)^{\alpha}+g^{\sigma\alpha}\left(k+p_{g_{2}}\right)^{\rho}-g^{\alpha\rho}\left(k+p_{g_{1}}\right)^{\sigma}.\end{equation}

The squared amplitude, after summing over colors and spins, are given
by\begin{eqnarray}
 &  & \sum_{\substack{{\rm color}\\
{\rm spin}}
}\left|\mathcal{M}(W^{+}\to t\bar{b}gg)\right|^{2}\nonumber \\
 & = & \frac{N_{C}^{2}-1}{2}\frac{N_{C}}{2}\left[\left|\mathcal{M}_{1}\right|^{2}+\left|\mathcal{M}_{2}\right|^{2}-\frac{1}{N_{C}^{2}}\left|\mathcal{M}_{1}+\mathcal{M}_{2}\right|^{2}\right],\label{eq:color-order-amps}\end{eqnarray}
where summing over the final state degrees of freedom is understood
in $\left|\mathcal{M}_{i}\right|^{2}$. The factorization of soft
and collinear singularities for color ordered amplitudes has been
discussed in the literature mainly for the leading color terms ($\mathcal{O}(N)$)\ \citep{Giele:1991vf,Giele:1993dj}.
For our calculation of the one cutoff phase space slicing method,
we will have to extend these results to the sub-leading color terms
($\mathcal{O}(1/N)$).

\subsection{Soft singularities}

The soft gluon behavior for an color ordered sub-amplitude is very
similar to the soft photon behavior of QED amplitudes. In QED, the
soft photon couples to a charged fermion line, resulting in an eikonal
factor multiplying the hard process. The key point is that only the
Feynman diagrams with soft photon coupled to the external charged
fermion lines will contribute in the soft photon limit. Technically,
no matter how many photons are radiated out from the charged fermion
dipole, the soft eikonal factor only knows about the external momentum
since there is no photon-photon interaction in QED theory. In QCD,
the soft pattern becomes much complicated due to the non-Abelian interaction.
However, it has been shown that the color ordered sub-amplitudes do
exhibit a factorization of the soft gluon singularities as in the
QED\ \citep{Bassetto:1984ik,Berends:1988zn}. This is because the
partons are ordered and form well defined color charge lines to which
the soft gluon can couple. The soft gluon behavior depends only on
the momenta of the external color charged lines to which the soft
gluon couples, and is independent of the number and type of other
partons in the process. Similarly, the soft factor is independent
of whether or not any color singlet particles such as electroweak
bosons are participating in the hard process.

\subsubsection{Factorization of the color ordered amplitudes in the soft limit}

In the limit of one gluon being soft, each term in the squared amplitudes
(cf. Eq.~\ref{eq:color-order-amps}) can be written as a factor multiplying
the Born level amplitude square. Since $t$ and $\bar{b}$ are the
external fermions which cannot be soft, there are only two sources
of the soft singularities: either from $g_{1}$ or from $g_{2}$.
When both of them are soft, it goes beyond the NLO ($\mathcal{O}(\alpha_{s}^{2})$)
and will not be considered in our calculation. 

\begin{figure}
\includegraphics[scale=0.5]{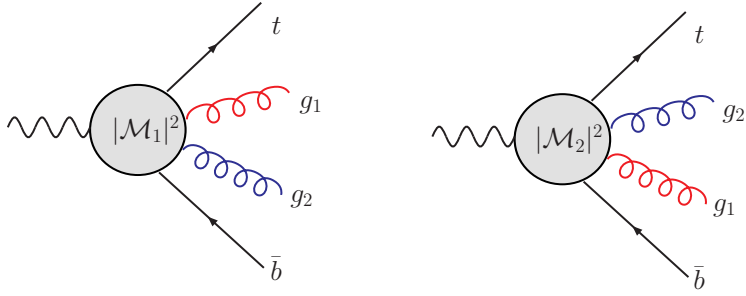}

\caption{Momentum label of the color ordered amplitudes.\label{fig:Momentum-label.}}

\end{figure}

First consider $\left|\mathcal{M}_{1,\,2}\right|^{2}$. The momentum
configurations which respect the color order are shown in Fig.~\ref{fig:Momentum-label.}.
In the soft gluon limit, $\left|\mathcal{M}_{1}\right|^{2}$ can be
factorized as follows:\begin{eqnarray}
\frac{N_{C}^{2}-1}{2}\frac{N_{C}}{2}\left|\mathcal{M}_{1}\right|^{2} & \xrightarrow{g_{1}\to0} & \frac{g_{s}^{2}N_{C}}{2}\, f\left(t,g_{1},g_{2}\right)\left|\mathcal{M}_{0}\right|^{2},\\
 & \xrightarrow{g_{2}\to0} & \frac{g_{s}^{2}N_{C}}{2}\, f(g_{1},g_{2},\bar{b})\left|\mathcal{M}_{0}\right|^{2},\end{eqnarray}
where the eikonal factor $f(a,s,b)$ is defined as\begin{equation}
f\left(a,s,b\right)=\frac{4s_{ab}}{s_{as}s_{sb}}-\frac{4m_{a}^{2}}{s_{as}^{2}}-\frac{4m_{b}^{2}}{s_{sb}^{2}}.\end{equation}
Here we define $s_{ij}=2p_{i}\cdot p_{j}$ for both massive and massless
partons. The color factor $(N_{C}^{2}-1)/2$ has been absorbed into
the Born level amplitude square. Similarly, $\left|\mathcal{M}_{2}\right|^{2}$
can be factorized in the soft gluon limit as follows:\begin{eqnarray}
\frac{N_{C}^{2}-1}{2}\frac{N_{C}}{2}\left|\mathcal{M}_{2}\right|^{2} & \xrightarrow{g_{1}\to0} & \frac{g_{s}^{2}N_{C}}{2}\, f\left(g_{2},g_{1},\bar{b}\right)\left|\mathcal{M}_{0}\right|^{2},\\
 & \xrightarrow{g_{2}\to0} & \frac{g_{s}^{2}N_{C}}{2}\, f(t,g_{2},g_{1})\left|\mathcal{M}_{0}\right|^{2}.\end{eqnarray}
Now let us consider the QED-like term $\left|\mathcal{M}_{1}+\mathcal{M}_{2}\right|^{2}$.
 This interference term exhibits the canonical eikonal factorization
in the soft limit. The soft gluon only knows about the external color
charged fermion lines, leading to the following factorization\begin{eqnarray}
-\frac{N_{C}^{2}-1}{2}\frac{1}{2N_{C}}\left|\mathcal{M}_{1}+\mathcal{M}_{2}\right|^{2} & \xrightarrow{g_{1}\to0} & -\frac{g_{s}^{2}}{2N_{C}}\, f\left(t,g_{1},\bar{b}\right)\left|\mathcal{M}_{0}\right|^{2},\\
 & \xrightarrow{g_{2}\to0} & -\frac{g_{s}^{2}}{2N_{C}}\, f(t,g_{2},\bar{b})\left|\mathcal{M}_{0}\right|^{2}.\end{eqnarray}

\subsubsection{soft singularities of $\left|\mathcal{M}_{1}\right|^{2}$ }

In order to slice the phase space we introduce the following Heaviside
functions $\Theta(i,j,k)$ and $\bar{\Theta}(i,j,k)$:\begin{eqnarray}
\Theta(i,j,k) & \equiv & \Theta(s_{ij}+s_{jk}-2s_{min}),\\
\bar{\Theta}(i,j,k) & \equiv & \Theta(2s_{min}-s_{ij}-s_{jk}).\end{eqnarray}
With the help of $\Theta$-functions we can split the phase space
of the real emission corrections into four regions:\begin{eqnarray}
1 & = & \left(\Theta_{tg_{1}g_{2}}+\bar{\Theta}_{tg_{1}g_{2}}\right)\left(\Theta_{g_{1}g_{2}\bar{b}}+\bar{\Theta}_{g_{1}g_{2}\bar{b}}\right)\nonumber \\
 & = & \Theta_{tg_{1}g_{2}}\Theta_{g_{1}g_{2}\bar{b}}+\Theta_{tg_{1}g_{2}}\bar{\Theta}_{g_{1}g_{2}\bar{b}}+\bar{\Theta}_{tg_{1}g_{2}}\Theta_{g_{1}g_{2}\bar{b}}-\bar{\Theta}_{tg_{1}g_{2}}\bar{\Theta}_{g_{1}g_{2}\bar{b}}.\label{eq:pss-condition}\end{eqnarray}
The first term denotes the region where two gluons are both hard (no
soft singularities), but one should keep in mind that there might
exist the collinear divergences, depending on the masses of external
fermions. The second term denotes the region where the gluon $g_{2}$
is soft but the gluon $g_{1}$ is hard. The third term denotes the
region where $g_{1}$ is soft and $g_{2}$ is hard. The last term
denotes the region where $g_{1}$ and $g_{2}$ are both soft. As discussed
before, this region only contributes when one calculates the quantum
corrections beyond the next-to-leading order and therefore is ignored
in this work. The negative sign ahead of the fourth term is to get
rid of the double counting from the regions described in the second
and third terms. 

When $g_{1}$ is soft, the eikonal factor $f(t,g_{1},g_{2})$,\begin{eqnarray}
f(t,g_{1},g_{2}) & = & \frac{4s_{tg_{2}}}{s_{tg_{1}}s_{g_{1}g_{2}}}-\frac{4m_{t}^{2}}{s_{tg_{1}}^{2}},\end{eqnarray}
can be calculated in the center frame of ($t,g_{2}$) system where
we choose the explicit momentum of final state particles as follows:\begin{eqnarray*}
p_{t} & = & \left(E_{t},0,0,\beta_{t}E_{t}\right),\\
p_{g_{2}} & = & \left(\beta_{t}E_{t},0,0,-\beta_{t}E_{t}\right),\\
p_{g_{1}} & = & \left(E_{g_{1}},0,E_{g_{1}}\sin\theta,E_{g_{1}}\cos\theta\right).\end{eqnarray*}
Here, $E_{t}$, $E_{g_{1}}$ and $E_{g_{2}}$ is the energy of the
top quark, the soft gluon ($g_{1}$) and the hard gluon ($g_{2}$),
respectively, and $\theta$ is the angle between the soft gluon $g_{1}$
and the top quark. It is easy to show that the eikonal factor can
be written as\begin{equation}
f(t,g_{1},g_{2})=\frac{1}{E_{g_{1}}^{2}E_{t}}\left(\frac{2\sqrt{s_{tg_{2}}+m_{t}^{2}}}{(1+\cos\theta)(1-\beta_{t}\cos\theta)}-\frac{m_{t}^{2}}{E_{t}(1-\beta_{t}\cos\theta)^{2}}\right),\label{eq:eikonal_factor_wtb}\end{equation}
where\begin{eqnarray*}
s_{tg_{2}} & = & 2\beta E_{t}^{2}(1+\beta),\\
\beta_{t} & = & \frac{s_{tg_{2}}}{s_{tg_{2}}+2m^{2}},\\
E_{t} & = & \frac{s_{tg_{2}}+2m_{t}^{2}}{2\sqrt{s_{tg_{2}}+m_{t}^{2}}}.\end{eqnarray*}
Now let us calculate the phase space boundary condition with the choice
of momentum above. The sum of $s_{tg_{1}}$ and $s_{g_{1}g_{2}}$
is given by\begin{equation}
s_{tg_{1}}+s_{g_{1}g_{2}}=2E_{t}E_{g_{1}}(1+\beta_{t})=2E_{g_{1}}\sqrt{s_{tg_{2}}+m_{t}^{2}}.\label{eq:tg1g2_soft_max}\end{equation}
Substituting Eq.~\ref{eq:tg1g2_soft_max} into the soft phase space
boundary condition, we obtain the upper limit of $E_{g_{1}}$,\[
E_{g_{1}}\leq{\displaystyle \frac{s_{min}}{\sqrt{s_{tg_{2}}+m_{t}^{2}}}}\equiv E_{g_{1}}^{max}.\]
In the soft limit, the phase space is also factorizes as\begin{equation}
d^{n}\Phi_{4}\xrightarrow{p_{g_{1}}\rightarrow0}d^{n}\Phi_{3}\frac{1}{8\pi^{2}}\frac{(4\pi)^{\epsilon}}{\Gamma(1-\epsilon)}E_{g_{1}}^{1-2\epsilon}\sin^{1-2\epsilon}\theta\, dE_{g_{1}}d\theta.\label{eq:soft-factor-phasespace}\end{equation}
Then the soft gluon contribution is given by\begin{eqnarray}
 &  & I^{soft}(t,g_{1},g_{2})\nonumber \\
 & = & \int\bar{\Theta}_{tg_{1}g_{2}}\left[\frac{g_{s}^{2}N_{C}}{2}\left(\mu_{d}^{2}\right)^{\epsilon}f(t,g_{1},g_{2})\frac{d^{3}g_{1}}{(2\pi)^{d-1}2E_{g_{1}}}\right]\nonumber \\
 & = & N_{C}\frac{g_{s}^{2}}{2}\frac{1}{8\pi^{2}}\frac{\left(4\pi\mu_{d}^{2}\right)^{\epsilon}}{\Gamma(1-\epsilon)}\int_{0}^{E_{g_{1}}^{max}}E_{g_{1}}^{-1-2\epsilon}{\rm d}E_{g_{1}}\nonumber \\
 & \times & \int_{0}^{\pi}\sin^{1-2\epsilon}\theta\frac{1}{E_{t}}\left[\frac{2\sqrt{s_{tg_{2}}+m_{t}^{2}}}{\left(1+\cos\theta\right)\left(1-\beta_{t}\cos\theta\right)}-\frac{m_{t}^{2}}{E_{t}\left(1-\beta_{t}\cos\theta\right)^{2}}\right]{\rm d}\theta.\label{eq:soft_tg1g2_first}\end{eqnarray}
Using the formula listed in Appendix~\ref{sec:Useful-formula} we
evaluate the integral in Eq.~\ref{eq:soft_tg1g2_first} and obtain
the soft factor $I^{soft}\left(t,g_{1},g_{2}\right)$ as following,\begin{eqnarray}
I^{soft}(t,g_{1},g_{2}) & = & N_{C}\frac{g_{s}^{2}}{16\pi^{2}}\frac{1}{\Gamma(1-\epsilon)}\left(\frac{4\pi\mu_{d}^{2}}{s_{min}}\right)^{\epsilon}\left(\frac{s_{min}}{s_{tg_{2}}+m_{t}^{2}}\right)^{-\epsilon}\nonumber \\
 & \times & \left\{ \frac{1}{\epsilon_{IR}^{2}}--\frac{1}{\epsilon_{IR}}\left[\ln\left(1+\frac{s_{tg_{2}}}{m_{t}^{2}}\right)+2\ln2-1\right]\right.\nonumber \\
 &  & -\frac{\pi^{2}}{6}+2\ln^{2}2-2\ln2+\left[2\ln2+\frac{s_{tg_{2}}+2m_{t}^{2}}{s_{tg_{2}}}\right]\ln\left(1+\frac{s_{tg_{2}}}{m_{t}^{2}}\right)\nonumber \\
 &  & \left.-\frac{1}{2}\ln^{2}\left(1+\frac{s_{tg_{2}}}{m_{t}^{2}}\right)-2{\rm Li_{2}}\left(\frac{s_{tg_{2}}}{s_{tg_{2}}+m_{t}^{2}}\right)\right\} .\label{eq:soft_factor_tg1g2}\end{eqnarray}
Similarly, one can calculate the soft singularity when $g_{2}$ is
soft, yielding\begin{eqnarray}
I^{soft}(g_{1},g_{2},\bar{b}) & = & \int\bar{\Theta}_{g_{1}g_{2}\bar{b}}\left[\frac{g_{s}^{2}N_{C}}{2}\left(\mu_{d}^{2}\right)^{\epsilon}f(g_{1},g_{2},\bar{b})\frac{d^{3}g_{2}}{(2\pi)^{d-1}2E_{g_{2}}}\right]\nonumber \\
 & = & N_{C}\frac{g_{s}^{2}}{16\pi^{2}}\frac{1}{\Gamma(1-\epsilon)}\left(\frac{4\pi\mu_{d}^{2}}{s_{min}}\right)\left(\frac{s_{min}}{s_{g_{1}\bar{b}}}\right)^{-\epsilon}\nonumber \\
 & \times & \left\{ \frac{2}{\epsilon_{IR}^{2}}-\frac{4\ln2}{\epsilon_{IR}}+4\ln^{2}2-\frac{\pi^{2}}{3}\right\} .\label{eq:soft_factor_g1g2b_massless}\end{eqnarray}
This result is for $m_{b}=0$ only. If $m_{b}\neq0$, then the soft
factor can be obtained from Eq.~\ref{eq:soft_factor_tg1g2},\begin{eqnarray}
I^{soft}(g_{1},g_{2},\bar{b}) & =N_{C} & \frac{g_{s}^{2}}{16\pi^{2}}\frac{1}{\Gamma(1-\epsilon)}\left(\frac{4\pi\mu_{d}^{2}}{s_{min}}\right)^{\epsilon}\left(\frac{s_{min}}{s_{g_{1}\bar{b}}+m_{b}^{2}}\right)^{-\epsilon}\nonumber \\
 & \times & \left\{ \frac{1}{\epsilon^{2}}-\frac{1}{\epsilon}\left[\ln\left(1+\frac{s_{g_{1}\bar{b}}}{m_{b}^{2}}\right)+2\ln2-1\right]\right.\nonumber \\
 &  & -\frac{\pi^{2}}{6}+2\ln^{2}2-2\ln2+\left[2\ln2+\frac{s_{g_{1}\bar{b}}+2m_{b}^{2}}{s_{g_{1}\bar{b}}}\right]\ln\left(1+\frac{s_{g_{1}\bar{b}}}{m_{b}^{2}}\right)\nonumber \\
 &  & \left.-\frac{1}{2}\ln^{2}\left(1+\frac{s_{g_{1}\bar{b}}}{m_{b}^{2}}\right)-2{\rm Li_{2}}\left(\frac{s_{g_{1}\bar{b}}}{s_{g_{1}\bar{b}}+m_{b}^{2}}\right)\right\} .\label{eq:soft_factor_g1g2b_massive}\end{eqnarray}

\subsubsection{soft singularities of  $\left|\mathcal{M}_{2}\right|^{2}$}

The phase space can be splitted into four regions:\begin{eqnarray}
1 & = & \left(\Theta_{tg_{2}g_{1}}+\bar{\Theta}_{tg_{2}g_{1}}\right)\left(\Theta_{g_{2}g_{1}\bar{b}}+\bar{\Theta}_{g_{2}g_{1}\bar{b}}\right)\nonumber \\
 & = & \Theta_{tg_{2}g_{1}}\Theta_{g_{2}g_{1}\bar{b}}+\Theta_{tg_{2}g_{1}}\bar{\Theta}_{g_{2}g_{1}\bar{b}}+\bar{\Theta}_{tg_{2}g_{1}}\Theta_{g_{2}g_{1}\bar{b}}-\bar{\Theta}_{tg_{2}g_{1}}\bar{\Theta}_{g_{2}g_{1}\bar{b}}.\end{eqnarray}
The soft singularities of $\left|\mathcal{M}_{2}\right|^{2}$ can
be derived from the results of $\left|\mathcal{M}_{1}\right|^{2}$
by making the substitution $g_{1}\leftrightarrow g_{2}.$

\subsubsection{soft singularities of $\left|\mathcal{M}_{1}+\mathcal{M}_{2}\right|^{2}$}

The phase space can be splitted into\begin{eqnarray}
1 & = & \left(\Theta_{tg_{1}\bar{b}}+\bar{\Theta}_{tg_{1}\bar{b}}\right)\left(\Theta_{tg_{2}\bar{b}}+\bar{\Theta}_{tg_{2}\bar{b}}\right)\nonumber \\
 & = & \Theta_{tg_{1}\bar{b}}\Theta_{tg_{2}\bar{b}}+\Theta_{tg_{1}\bar{b}}\bar{\Theta}_{tg_{2}\bar{b}}+\bar{\Theta}_{tg_{1}\bar{b}}\Theta_{tg_{2}\bar{b}}-\bar{\Theta}_{tg_{1}\bar{b}}\bar{\Theta}_{tg_{2}\bar{b}}.\end{eqnarray}
The first term corresponds to the two hard gluons, the second term
denotes $E_{g_{1}}\to0$, the third term denotes $E_{g_{2}}\to0$
and the fourth term denotes both $E_{g_{1}}\to0$ and $E_{g_{2}}\to0$.
Since the eikonal factor is only determined by the two external fermion
lines ($t$ and $\bar{b}$), the soft factors of $g_{1}$ and $g_{2}$
should be the same, i.e.\begin{equation}
I^{soft}(t,g_{1},\bar{b})=I^{soft}(t,g_{2},\bar{b}).\end{equation}
The soft factor can easily be derived from that of $\bar{\Theta}_{tg_{1}g_{2}}\left|\mathcal{M}_{1}\right|^{2}$
by making the substitution $g_{2}\to\bar{b}$, because the soft gluon
contributions only know about the kinematics. It gives rise to the
soft factor as following,\begin{eqnarray}
I^{soft}(t,g_{1},\bar{b}) & = & \left.I^{soft}(t,g_{1},g_{2})\right|_{g_{2}\to\bar{b}}\nonumber \\
 & = & \left(-\frac{1}{N_{C}}\right)\frac{g_{s}^{2}}{16\pi^{2}}\frac{1}{\Gamma(1-\epsilon)}\left(\frac{4\pi\mu_{d}^{2}}{s_{min}}\right)^{\epsilon}\left(\frac{s_{min}}{s_{t\bar{b}}+m_{t}^{2}}\right)^{-\epsilon}\nonumber \\
 & \times & \left\{ \frac{1}{\epsilon_{IR}^{2}}--\frac{1}{\epsilon_{IR}}\left[\ln\left(1+\frac{s_{t\bar{b}}}{m_{t}^{2}}\right)+2\ln2-1\right]\right.\nonumber \\
 &  & -\frac{\pi^{2}}{6}+2\ln^{2}2-2\ln2+\left[2\ln2+\frac{s_{t\bar{b}}+2m_{t}^{2}}{s_{t\bar{b}}}\right]\ln\left(1+\frac{s_{t\bar{b}}}{m_{t}^{2}}\right)\nonumber \\
 &  & \left.-\frac{1}{2}\ln^{2}\left(1+\frac{s_{t\bar{b}}}{m_{t}^{2}}\right)-2{\rm Li_{2}}\left(\frac{s_{t\bar{b}}}{s_{t\bar{b}}+m_{t}^{2}}\right)\right\} .\label{eq:xxx}\end{eqnarray}

\subsection{Collinear singularities}

As mentioned above, there still exists collinear divergence even for
a hard gluon radiation. Below we will further slice the collinear
region of the phase space to derive the collinear divergences.

\subsubsection{Collinear singularities of $\left|\mathcal{M}\right|_{1}^{2}$}

We further slice the hard phase space of $\left|\mathcal{M}\right|_{1}^{2}$
as\begin{equation}
\Theta_{tg_{1}g_{2}}\Theta_{g_{1}g_{2}\bar{b}}\left|\mathcal{M}_{1}\right|^{2}=\left\{ \Theta_{tg_{1}g_{2}}\Theta_{g_{1}g_{2}\bar{b}}-C_{\mathcal{M}_{1}}+C_{\mathcal{M}_{1}}\right\} \left|\mathcal{M}_{1}\right|^{2},\end{equation}
where\begin{eqnarray}
C_{\mathcal{M}_{1}} & = & \Theta\left(s_{tg_{1}}-2s_{min}\right)\Theta\left(s_{g_{2}\bar{b}}-2s_{min}\right)\Theta\left(s_{min}-s_{g_{1}g_{2}}\right)\nonumber \\
 & + & \Theta\left(s_{t\bar{b}}-2s_{min}\right)\Theta\left(s_{g_{1}g_{2}}-2s_{min}\right)\Theta\left(s_{min}-s_{g_{2}\bar{b}}\right).\end{eqnarray}
The first term in $C_{\mathcal{M}_{1}}$ represents the phase space
region that neither $g_{1}$ or $g_{2}$ is not soft but $g_{1}\parallel g_{2}$,
while the second term denotes the phase space region of $g_{2}\parallel\bar{b}$. 

In the collinear region of $g_{1}\parallel g_{2}$, the color ordered
matrix element square exhibits the following factorization property,\begin{equation}
\frac{N_{C}}{2}\frac{N_{C}^{2}-1}{2}\left|\mathcal{M}_{1}\right|^{2}\xrightarrow{g_{1}\parallel g_{2}}\frac{g_{s}^{2}N_{C}}{2}f^{gg\to g}\left|\mathcal{M}_{0}\right|^{2},\end{equation}
where\begin{equation}
f^{gg\to g}\left(\xi\right)=\frac{2}{s_{g_{1}g_{2}}}\frac{1+\xi^{4}+\left(1-\xi\right)^{4}}{\xi\left(1-\xi\right)}-\frac{4m_{t}^{2}}{\xi^{2}s_{th}^{2}},\end{equation}
with the kinematics defined as\begin{equation}
p_{h}=p_{g_{1}}+p_{g_{2}},\quad p_{g_{1}}=\xi p_{h},\quad p_{g_{2}}=\left(1-\xi\right)p_{h}.\end{equation}
The phase space integration in the collinear region can also be written
in the following factorization form\begin{eqnarray}
 &  & \frac{d^{n}p_{g_{1}}}{\left(2\pi\right)^{n-1}2E_{g_{1}}}\frac{d^{n}p_{g_{2}}}{\left(2\pi\right)^{n-1}2E_{g_{2}}}\nonumber \\
 & = & \frac{1}{16\pi^{2}}\frac{\left(4\pi\right)^{\epsilon}}{\Gamma\left(1-\epsilon\right)}\left[s_{g_{1}g_{2}}\xi\left(1-\xi\right)\right]^{-\epsilon}ds_{g_{1}g_{2}}d\xi\times\frac{d^{d}p_{h}}{\left(2\pi\right)^{d-1}2E_{h}}.\end{eqnarray}
Bearing in mind that the integration is limited by the conditions
$s_{g_{1}g_{2}}<s_{min}$, $s_{tg_{1}}>2s_{min}$ and $s_{g_{2}\bar{b}}>2s_{min}$,
we integrate the matrix element over the collinear phase space and
obtain the following collinear factor,\begin{eqnarray}
C_{\mathcal{M}_{1}}^{gg\to g} & = & \frac{\alpha_{s}}{4\pi}\frac{N_{C}}{\Gamma\left(1-\epsilon\right)}\left(\frac{4\pi\mu}{s_{min}}\right)^{2}\biggl\{\frac{2}{\epsilon_{IR}}\left[\ln\frac{2s_{min}}{s_{tg_{1}}}+\ln\frac{2s_{min}}{s_{g_{1}\bar{b}}}+\frac{11}{6}\right]\nonumber \\
 &  & \qquad\qquad-\frac{2m_{t}^{2}}{s_{tg_{1}}}-\frac{2\pi^{2}}{3}+\frac{67}{9}-\ln^{2}\frac{2s_{min}}{s_{tg_{1}}}-\ln^{2}\frac{2s_{min}}{s_{g_{1}\bar{b}}}\biggr\}.\end{eqnarray}

In the collinear region of $g_{2}\parallel\bar{b}$, the color ordered
matrix element square exhibits the following factorization property,\begin{equation}
\frac{N_{C}}{2}\frac{N_{C}^{2}-1}{2}\left|\mathcal{M}_{1}\right|^{2}\xrightarrow{g_{2}\parallel\bar{b}}\frac{g_{s}^{2}N_{C}}{2}f^{qg\to q}\left|\mathcal{M}_{0}\right|^{2},\label{eq:msq-coll}\end{equation}
where \begin{equation}
f^{qg\to q}\left(\xi\right)=\frac{2}{s_{g_{2}\bar{b}}}\frac{1+\xi^{2}-\epsilon\left(1-\xi\right)^{2}}{\left(1-\xi\right)}.\end{equation}
Again, after integrating over the collinear phase space, we obtain
the following collinear factor,\begin{equation}
C_{\mathcal{M}_{1}}^{qg\to q}=\frac{\alpha_{s}}{4\pi}\frac{N_{C}}{\Gamma\left(1-\epsilon\right)}\left(\frac{4\pi\mu}{s_{min}}\right)^{2}\biggl\{\frac{2}{\epsilon_{IR}}\left[\ln\frac{2s_{min}}{s_{g_{1}\bar{b}}}+\frac{3}{4}\right]-\frac{\pi^{2}}{3}+\frac{7}{2}-\ln^{2}\frac{2s_{min}}{s_{g_{1}\bar{b}}}\biggr\}.\end{equation}

\subsubsection{Collinear singularities of $\left|\mathcal{M}_{2}\right|^{2}$}

Similar to the case of $\left|\mathcal{M}_{1}\right|^{2}$, we split
the hard phase space of $\left|\mathcal{M}_{2}\right|^{2}$ as\begin{equation}
\Theta_{tg_{2}g_{1}}\Theta_{g_{2}g_{1}\bar{b}}\left|\mathcal{M}_{2}\right|^{2}=\biggl[\Theta_{tg_{2}g_{1}}\Theta_{g_{2}g_{1}\bar{b}}-C_{\mathcal{M}_{2}}+C_{\mathcal{M}_{2}}\biggr]\left|\mathcal{M}_{2}\right|^{2},\end{equation}
where\begin{eqnarray}
C_{\mathcal{M}_{2}} & = & \Theta\left(s_{tg_{2}}-2s_{min}\right)\Theta\left(s_{g_{1}\bar{b}}-2s_{min}\right)\Theta\left(s_{min}-s_{g_{2}g_{1}}\right)\nonumber \\
 & + & \Theta\left(s_{t\bar{b}}-2s_{min}\right)\Theta\left(s_{g_{2}g_{1}}-2s_{min}\right)\Theta\left(s_{min}-s_{g_{1}\bar{b}}\right).\end{eqnarray}
The first (second) term in $C_{\mathcal{M}_{2}}$ represents the collinear
region of $g_{2}\parallel g_{1}$ ($g_{1}\parallel\bar{b}$), respectively.
The collinear factors can be derived on the analogy of those of $\left|\mathcal{M}_{1}\right|^{2}$,\begin{equation}
C_{\mathcal{M}_{2}}^{gg\to g}=C_{\mathcal{M}_{1}}^{gg\to g},\quad C_{\mathcal{M}_{2}}^{qg\to q}=C_{\mathcal{M}_{2}}^{qg\to q}.\end{equation}

\subsubsection{Collinear singularities of $\left|\mathcal{M}_{1}+\mathcal{M}_{2}\right|^{2}$}

We split the hard phase space of $\left|\mathcal{M}_{1}+\mathcal{M}_{2}\right|^{2}\equiv\mathcal{M}_{1+2}$
as\begin{equation}
\Theta_{tg_{1}\bar{b}}\Theta_{tg_{2}\bar{b}}\left|\mathcal{M}_{1+2}\right|^{2}=\biggl[\Theta_{tg_{1}\bar{b}}\Theta_{tg_{2}\bar{b}}-C_{\mathcal{M}_{1+2}}+C_{\mathcal{M}_{1+2}}\biggr]\left|\mathcal{M}_{1+2}\right|^{2},\end{equation}
where\begin{eqnarray}
C_{\mathcal{M}_{1+2}} & = & \Theta\left(s_{tg_{1}}-2s_{min}\right)\Theta\left(s_{g_{2}\bar{b}}-2s_{min}\right)\Theta\left(s_{min}-s_{g_{1}\bar{b}}\right)\nonumber \\
 & + & \Theta\left(s_{tg_{2}}-2s_{min}\right)\Theta\left(s_{g_{1}\bar{b}}-2s_{min}\right)\Theta\left(s_{min}-s_{g_{2}\bar{b}}\right).\end{eqnarray}
The first (second) term in $C_{\mathcal{M}_{1+2}}$ represents the
collinear region of $g_{1}\parallel\bar{b}$ ($g_{2}\parallel\bar{b}$),
respectively. In both collinear regions, the matrix element square
can be factorized as\begin{equation}
-\frac{N_{C}^{2}-1}{2}\frac{1}{2N_{C}}\left|\mathcal{M}_{1+2}\right|^{2}\xrightarrow{g_{i}\parallel\bar{b}}-\frac{g_{s}^{2}}{2N_{C}}f^{qg_{i}\to q}\left|\mathcal{M}_{0}\right|^{2}.\end{equation}
Integration over the collinear region gives rise to the following
collinear factor,\begin{eqnarray}
 &  & C_{\mathcal{M}_{1+2}}^{qg_{1}\to q}=C_{\mathcal{M}_{1+2}}^{qg_{2}\to q}\nonumber \\
 & = & -\frac{\alpha_{s}}{4\pi}\frac{1}{N_{C}}\frac{1}{\Gamma\left(1-\epsilon\right)}\left(\frac{4\pi\mu^{2}}{s_{min}}\right)^{2}\biggl\{\frac{2}{\epsilon_{IR}}\left[\ln\frac{2s_{min}}{s_{t\bar{b}}}+\frac{3}{4}\right]-\frac{2m_{t}^{2}}{s_{t\bar{b}}}-\frac{\pi^{2}}{3}+\frac{7}{2}-\ln^{2}\frac{2s_{min}}{s_{t\bar{b}}}\biggr\}.\end{eqnarray}

\subsubsection{Collinear singularities of $W\to t\bar{b}q\bar{q}$}

Now let us consider the last piece of the real emission correction,
$W^{+}\to t\bar{b}q\bar{q}$, which only involves the collinear divergence.
We split the phase space into two parts,\begin{equation}
\Theta\left(s_{q\bar{q}}-s_{min}\right)+\Theta\left(s_{min}-s_{q\bar{q}}\right),\end{equation}
where the first (second) term represents the finite (collinear) region,
respectively. In the collinear region $q\parallel\bar{q}$, the matrix
element square can factorized as\begin{equation}
\left|\mathcal{M}\left(t\bar{b}q\bar{q}\right)\right|^{2}\xrightarrow{q\parallel\bar{q}}\frac{g_{s}}{2}f^{q\bar{q}\to g}\left|\mathcal{M}_{0}\right|^{2},\end{equation}
where\begin{equation}
f^{q\bar{q}\to g}\left(\xi\right)=\frac{2}{s_{q\bar{q}}}\frac{\xi^{2}+\left(1-\xi\right)^{2}-\epsilon}{1-\epsilon}.\end{equation}
Integration over the collinear region gives rise to the following
collinear factor\begin{equation}
C^{q\bar{q}\to g}=\frac{\alpha_{s}}{4\pi}\frac{1}{\Gamma\left(1-\epsilon\right)}\left(\frac{4\pi\mu^{2}}{s_{min}}\right)^{\epsilon}\left\{ -\frac{2}{3\epsilon_{IR}}-\frac{10}{9}\right\} \times n_{f},\end{equation}
where $n_{f}$ denotes the light quark flavors.

\subsection{complete result of the IR singularities of the real emission corrections}

The complete soft and collinear singularities of the real emission
corrections to the process $gb\to tW$ are given by\begin{eqnarray}
\hat{\sigma}_{IR} & = & \left(I_{S}+I_{C}\right)\left|\mathcal{M}_{0}\right|^{2},\end{eqnarray}
where the soft factor $I_{S}$ is given by\begin{eqnarray}
I_{S} & = & \frac{1}{2}\biggl[I^{soft}\left(t,g_{1},\bar{b}\right)+I^{soft}\left(t,g_{2},\bar{b}\right)+I^{soft}\left(t,g_{2},g_{1}\right)\nonumber \\
 &  & \quad+I^{soft}\left(g_{2},g_{1},\bar{b}\right)+I^{soft}\left(t,g_{1},\bar{b}\right)+I^{soft}\left(t,g_{2},\bar{b}\right)\biggr],\end{eqnarray}
while the collinear factor $I_{C}$ is given by\begin{eqnarray}
I_{C} & = & \frac{1}{2}\biggl[C_{\mathcal{M}_{1}}^{gg\to g}+C_{\mathcal{M}_{1}}^{qg\to q}+C_{\mathcal{M}_{2}}^{gg\to g}+C_{\mathcal{M}_{2}}^{qg\to q}+C_{\mathcal{M}_{1+2}}^{qg_{1}\to q}+C_{\mathcal{M}_{1+2}}^{qg_{2}\to q}\biggr]+C^{q\bar{q}\to g}.\end{eqnarray}
The factor {}``$2$'' in above equations accounts for the two identical
gluons in the final state. After summing over both soft and collinear
factors, we obtain the IR divergences of the real emission corrections
as following\begin{eqnarray}
\hat{\sigma}_{IR} & = & \frac{\alpha_{s}}{4\pi}C_{\epsilon}\biggl\{\frac{26}{3\epsilon_{IR}^{2}}-\frac{6}{\epsilon_{IR}}\ln\frac{s_{tg}}{m_{t}^{2}}-\frac{6}{\epsilon_{IR}}\ln\frac{s_{g\bar{b}}}{m_{t}^{2}}+\frac{2}{\epsilon_{IR}}\ln\frac{s_{t\bar{b}}}{m_{t}^{2}}\nonumber \\
 &  & \qquad\quad+\frac{1}{\epsilon_{IR}}\left(11-\frac{2}{3}n_{f}\right)+\frac{20}{3\epsilon_{IR}}+{\rm finite}\biggr\}\left|\mathcal{M}_{0}\right|^{2}.\end{eqnarray}
It is clear that the IR divergences in the real emission corrections
exactly cancel the UV divergences in the virtual corrections after
we cross the gluon and bottom quark into the initial state and the
$W$-boson to the final state, i.e.\begin{equation}
s_{tg}\to-t_{1},\qquad s_{g\bar{b}}\to s,\qquad s_{t\bar{b}}\to m_{t}^{2}-m_{W}^{2}-u.\end{equation}

\section{Resolved real emission corrections}

The real emission corrections in the resolved phase phase region are
finite and thus can be calculated numerically in 4 dimensions, using
the canonical Monte Carlo method. With the implement of the phase
space slicing conditions described above, the integration will depend
upon the input value of $s_{min}$. Since the cutoff $s_{min}$ is
introduced in the calculation only for a technical reason and is unrelated
to any physical quantity, the inclusive rate must not depend on it.
In other words, the sum of all contributions, virtual, resolved, and
unresolved corrections must be independent of $s_{min}$. This is
the case as long as $s_{min}$ is small enough so that the soft and
collinear approximations are valid. However, numerical cancellation
in the Monte Carlo integration becomes unstable if $s_{min}$ is too
small. Furthermore, the jet-finding algorithm and other infrared-safe
experimental observables should also be defined in a way such that
they are consistent with the choice of $s_{min}$. In practice, one
wants to choose the largest $s_{min}$ possible within these constraints
in order to minimize the processing time of the Monte Carlo integration
program. Since the phase space slicing conditions highly depend on
the kinematics of the final state particles, it is better to calculate
the resolved real emission corrections term by term in Eq.\ \ref{eq:color-order-amps}
to achieve better cancellation of the $s_{min}$ dependence.

\section{Conclusion}

In this paper we present a detailed calculation of the NLO QCD corrections
to the $tW$ associated production using the one cutoff phase space
slicing method. The corrections to the this process have been calculated
independently by two groups\ \citep{Zhu:2002uj,Campbell:2005bb},
a number of differences were found. Therefore, a third party calculation
is needed to make a crossing check. Furthermore, the fully analytic
expression is still missing in the literature. In this note, we first
calculated the virtual corrections using the dimensional regularization.
The final results are shown completely in terms of the Passarino-Veltman
scalar functions. We then use the one cutoff phase space slicing method
to calculate the infrared singularities of the real emission corrections.
The soft and collinear factorizations of the color ordered amplitudes
are shown in details. The phenomenology study will be presented in
the forthcoming paper.

\begin{acknowledgments}
We gratefully acknowledge stimulating discussions with Chuan-Ren Chen
and C.-P. Yuan. This work is supported in part by the U.~S.~Department
of Energy under Grant No.~DE-FG03-94ER40837.
\end{acknowledgments}
\appendix

\section{Useful formula\label{sec:Useful-formula}}

The integration over the soft and collinear regions can be easily
evaluated. Below we list the useful formula in our calculation\ \citep{Beenakker:1988bq}.
All the integrals we need are of the form\begin{equation}
I^{(k,l)}=\int_{0}^{\pi}{\rm d}\theta_{1}\,\frac{\sin^{1-2\epsilon}\theta_{1}}{(a+b\cos\theta_{1})^{k}}\int_{0}^{\pi}{\rm d}\theta_{2}\frac{\sin^{-2\epsilon}\theta_{2}}{(A+B\cos\theta_{1}+C\sin\theta_{1}\cos\theta_{2})^{l}}.\label{eq:speical_function}\end{equation}
The first point to notice is that there are four classes of integrals,
depending on the collinear structure. The $\left[ab\right]$ and $\left[AB\right]$
variable can both be either {}``collinear'' or not, which yields
four combinations. In the case of the $\left[ab\right]$ variable,
collinear divergence appears when $a^{2}=b^{2}$ so that $(a+b\cos\theta_{1})\to a(1\pm\cos\theta_{1}).$
Then the zero occurs at the edge of the $\theta_{1}$ integration
region and is not integrable. For the $\left[ABC\right]$ variable
the same comments apply for $\theta_{2}$ when $A^{2}=B^{2}+C^{2}$.
For our needs, the soft eikonal factor does not involve the angle
$\theta_{2}$, therefore the integration becomes more simpler.

When $A^{2}\neq B^{2}+C^{2}$, and $b=-a$, we use \begin{eqnarray}
I^{(1,1)} & = & \frac{\pi}{a(A+B)}\left\{ -\frac{1}{\epsilon}+\ln\left[\frac{(A+B)^{2}}{A^{2}-B^{2}-C^{2}}\right]\right.\nonumber \\
 &  & -\epsilon\left[\ln^{2}\left(\frac{A-\sqrt{B^{2}+C^{2}}}{A+B}\right)-\frac{1}{2}\ln^{2}\left(\frac{A+\sqrt{B^{2}+C^{2}}}{A-\sqrt{B^{2}+C^{2}}}\right)\right.\nonumber \\
 &  & \left.\left.\,\,\,\,\,\,+2{\rm Li}_{2}\left(-\frac{B+\sqrt{B^{2}+C^{2}}}{A-\sqrt{B^{2}+C^{2}}}\right)-2{\rm Li}_{2}\left(\frac{B-\sqrt{B^{2}+C^{2}}}{A+B}\right)\right]\right\} +\mathcal{O}(\epsilon^{2}),\label{eq:special_function_1}\end{eqnarray}
While when $b\neq-a$ we use\begin{eqnarray}
I^{(0,1)} & = & \frac{\pi}{\sqrt{B^{2}+C^{2}}}\left\{ \ln\left(\frac{A+\sqrt{B^{2}+C^{2}}}{A-\sqrt{B^{2}+C^{2}}}\right)\right.\nonumber \\
 &  & +2\epsilon\left[{\rm Li}_{2}\left(\frac{2\sqrt{B^{2}+C^{2}}}{A+\sqrt{B^{2}+C^{2}}}\right)+\frac{1}{4}\ln^{2}\left(\frac{A+\sqrt{B^{2}+C^{2}}}{A-\sqrt{B^{2}+C^{2}}}\right)\right]+\mathcal{O}(\epsilon^{2}),\label{eq:special_fucntion_2}\\
I^{(0,2)} & = & \frac{2\pi}{A^{2}-B^{2}-C^{2}}\left[1+\epsilon\frac{A}{\sqrt{B^{2}+C^{2}}}\ln\left(\frac{A+\sqrt{B^{2}+C^{2}}}{A-\sqrt{B^{2}+C^{2}}}\right)\right].\label{eq:special_fucntion_3}\end{eqnarray}
Finally, when $A^{%2
}=B^{2}+C^{2}$, and $b=-a$, we have\begin{equation}
I^{(1,1)}=-\frac{1}{\epsilon}\frac{\pi}{aA}\left(\frac{A+B}{2A}\right)^{-1-\epsilon}\left[1+\epsilon^{2}{\rm Li}_{2}\left(\frac{A-B}{2A}\right)\right].\label{eq:speical_function_4}\end{equation}

\section{Passiano-Veltman scalar functions\label{sec:scalarfunction}}

Here we list out the scalar functions used in this calculation. The
general scalar functions are defined as follows:\begin{eqnarray}
 &  & A_{0}\left(m^{2}\right)=\mu^{2\epsilon}\int\frac{d^{n}q}{\left(2\pi\right)^{n}}\frac{1}{q^{2}-m^{2}},\\
 &  & B_{0}\left(p_{1}^{2};m_{1}^{2},m_{2}^{2}\right)=\mu^{2\epsilon}\int\frac{d^{n}q}{\left(2\pi\right)^{n}}\frac{1}{\left[q^{2}-m_{1}^{2}\right]\left[\left(q+p_{1}\right)^{2}-m_{2}^{2}\right]},\\
 &  & C_{0}\left(p_{1}^{2},p_{2}^{2},p_{3}^{2};m_{1}^{2},m_{2}^{2},m_{3}^{2}\right)\nonumber \\
 &  & =\mu^{2\epsilon}\int\frac{d^{n}q}{\left(2\pi\right)^{n}}\frac{1}{\left[q^{2}-m_{1}^{2}\right]\left[\left(q+p_{1}\right)^{2}-m_{2}^{2}\right]\left[\left(q+p_{1}+p_{2}\right)^{2}-m_{3}^{2}\right]},\\
 &  & D_{0}\left(p_{1}^{2},p_{2}^{2},p_{3}^{2},p_{4}^{2};s_{12},s_{23};m_{1}^{2},m_{2}^{2},m_{3}^{2},m_{4}^{2}\right)\\
 &  & =\mu^{2\epsilon}\int\frac{d^{n}q}{\left(2\pi\right)^{n}}\frac{1}{\left[q^{2}-m_{1}^{2}\right]\left[\left(q+p_{1}\right)^{2}-m_{2}^{2}\right]\left[\left(q+p_{1}+p_{2}\right)^{2}-m_{3}^{2}\right]\left[\left(q+p_{1}+p_{2}+p_{3}\right)^{2}-m_{4}^{2}\right]},\nonumber \end{eqnarray}
where $n=4-2\epsilon$, $\mu$ is the scale introduced so that the
integrals preserve their natural dimensions. We follow the notation
of Ref.\ \citep{Ellis:2007qk} to calculate the scalar integrals
in the spacelike region, $p_{i}^{2}<0$ and $s_{ij}\equiv\left(p_{i}+p_{j}\right)^{2}<0$.
The analytic continuation is performed by restoring the $i\varepsilon$,\[
p_{i}^{2}\to p_{i}^{2}+i\varepsilon,\quad s_{ij}\to s_{ij}+i\varepsilon,\quad m_{i}^{2}\to m_{i}^{2}-i\varepsilon.\]

The divergent three-point scalar functions used in this work are given
as follows:\begin{eqnarray}
 &  & C_{0}^{V_{i}}\left(0,0,p_{3}^{2};0,0,0\right)\nonumber \\
 &  & =\frac{1}{16\pi^{2}}C_{\epsilon}\biggl\{\frac{1}{\epsilon^{2}}-\frac{1}{\epsilon}\ln\left(\frac{-p_{3}^{2}}{m_{t}^{2}}\right)+\frac{1}{2}\ln^{2}\left(\frac{-p_{3}^{2}}{m_{t}^{2}}\right)-\frac{1}{6}\pi^{2}\biggr\},\label{eq:c1}\\
 &  & C_{0}^{V_{i}}\left(0,p_{2}^{2},m_{t}^{2};0,0,m_{t}^{2}\right)\nonumber \\
 &  & =\frac{1}{16\pi^{2}}C_{\epsilon}\biggl\{\frac{1}{2\epsilon^{2}}+\frac{1}{\epsilon}\ln\left(\frac{m_{t}^{2}}{m_{t}^{2}-p_{2}^{2}}\right)+\frac{1}{2}\ln^{2}\left(\frac{m_{t}^{2}}{m_{t}^{2}-p_{2}^{2}}\right)-{\rm Li}_{2}\left(\frac{-p_{2}^{2}}{m_{t}^{2}-p_{2}^{2}}\right)\biggr\},\label{eq:c2}\\
 &  & C_{0}^{V_{i}}\left(0,p_{2}^{2},p_{3}^{2};0,0,m_{t}^{2}\right)\nonumber \\
 &  & =\frac{1}{16\pi}C_{\epsilon}\frac{1}{p_{2}^{2}-p_{3}^{2}}\biggl\{\frac{1}{\epsilon}\ln\left(\frac{m_{t}^{2}-p_{3}^{2}}{m_{t}^{2}-p_{2}^{2}}\right)+{\rm Li}_{2}\left(\frac{p_{2}^{2}}{m_{t}^{2}}\right)-{\rm Li}_{2}\left(\frac{p_{3}^{2}}{m_{t}^{2}}\right)\nonumber \\
 &  & \qquad\qquad\qquad+\ln^{2}\left(\frac{m_{t}^{2}-p_{2}^{2}}{m_{t}^{2}}\right)-\ln^{2}\left(\frac{m_{t}^{2}-p_{3}^{2}}{m_{t}^{2}}\right)\biggr\}.\label{eq:c3}\end{eqnarray}
where \[
C_{\epsilon}\equiv\left(\frac{4\pi\mu^{2}}{m_{t}^{2}}\right)^{\epsilon}\Gamma\left(1+\epsilon\right).\]

The divergent four-point scalar functions are given as follows:\begin{eqnarray}
 &  & D_{0}^{B_{1}}\left(0,m_{t}^{2},0,m_{W}^{2};u,t;0,0,m_{t}^{2},m_{t}^{2}\right)\nonumber \\
 &  & =\frac{1}{16\pi^{2}}C_{\epsilon}\frac{1}{\left(u-m_{t}^{2}\right)\left(t-m_{t}^{2}\right)}\biggl\{\frac{1}{2\epsilon^{2}}-\frac{1}{\epsilon}\ln\left[\frac{\left(m_{t}^{2}-u\right)\left(m_{t}^{2}-t\right)}{\left(m_{t}^{2}-m_{W}^{2}\right)m_{t}^{2}}\right]-\frac{\pi^{2}}{6}\nonumber \\
 &  & \qquad\qquad-\ln^{2}\left(\frac{m_{t}^{2}-m_{W}^{2}}{m_{t}^{2}}\right)+2\ln\left(\frac{m_{t}^{2}-t}{m_{t}^{2}-m_{W}^{2}}\right)\ln\left(\frac{m_{t}^{2}-u}{m_{t}^{2}}\right)\nonumber \\
 &  & \qquad\qquad-2{\rm Li}_{2}\left(1-\frac{m_{t}^{2}-m_{W}^{2}}{m_{t}^{2}-t}\right)-2{\rm Li}_{2}\left(1-\frac{m_{t}^{2}-m_{W}^{2}}{m_{t}^{2}-u}\right)\biggr\},\\
 &  & D_{0}^{B_{2}}\left(0,0,m_{t}^{2},m_{W}^{2};s,u;0,m_{t}^{2},0,0\right)\nonumber \\
 &  & =\frac{1}{16\pi^{2}}C_{\epsilon}\frac{1}{s\left(u-m_{t}^{2}\right)}\biggl\{\frac{3}{2}\frac{1}{\epsilon^{2}}-\frac{1}{\epsilon}\left[2\ln\left(1-\frac{u}{m_{t}^{2}}\right)+\ln\left(\frac{s}{m_{t}^{2}}\right)-\ln\left(1-\frac{m_{W}^{2}}{m_{t}^{2}}\right)\right]\nonumber \\
 &  & \quad-2{\rm Li}_{2}\left(\frac{u-m_{W}^{2}}{u-m_{t}^{2}}\right)+2\ln\left(\frac{-s}{m_{t}^{2}}\right)\ln\left(1-\frac{u}{m_{t}^{2}}\right)-\ln^{2}\left(1-\frac{m_{W}^{2}}{m_{t}^{2}}\right)-\frac{2}{3}\pi^{2}\biggr\},\label{eq:d2}\\
 &  & D_{0}^{B_{3}}=D_{0}^{B_{2}}\biggr|_{u\to t}.\label{eq:d3}\end{eqnarray}

The scalar functions exhibit certain symmetries under interchange
their arguments (either rotation or inversion). The symmetry properties
of the triangle- and box-loop integrals used in this calculation are
listed as follows\ \citep{Ellis:2007qk}:\\
(1) $C_{0}$-function\begin{eqnarray}
C_{0}\left(p_{1}^{2},p_{2}^{2},p_{3}^{2};m_{1}^{2},m_{2}^{2},m_{3}^{2}\right) & = & C_{0}\left(p_{2}^{2},p_{3}^{2},p_{1}^{2};m_{2}^{2},m_{3}^{2},m_{1}^{2}\right),\\
C_{0}\left(p_{1}^{2},p_{2}^{2},p_{3}^{2};m_{1}^{2},m_{2}^{2},m_{3}^{2}\right) & = & C_{0}\left(p_{3}^{2},p_{2}^{2},p_{1}^{2};m_{3}^{2},m_{2}^{2},m_{1}^{2}\right).\end{eqnarray}
(2) $D_{0}$-function\begin{eqnarray}
 &  & D_{0}\left(p_{1}^{2},p_{2}^{2},p_{3}^{2},p_{4}^{2};s_{12},s_{23};m_{1}^{2},m_{2}^{2},m_{3}^{2},m_{4}^{2}\right)\nonumber \\
 & = & D_{0}\left(p_{2}^{2},p_{3}^{2},p_{4}^{2},p_{1}^{2};s_{23},s_{12};m_{2}^{2},m_{3}^{2},m_{4}^{2},m_{1}^{2}\right),\\
\nonumber \\ &  & D_{0}\left(p_{1}^{2},p_{2}^{2},p_{3}^{2},p_{4}^{2};s_{12},s_{23};m_{1}^{2},m_{2}^{2},m_{3}^{2},m_{4}^{2}\right)\nonumber \\
 & = & D_{0}\left(p_{4}^{2},p_{3}^{2},p_{2}^{2},p_{1}^{2};s_{12},s_{23};m_{1}^{2},m_{4}^{2},m_{3}^{2},m_{2}^{2}\right).\end{eqnarray}

\section{Form factors of the virtual corrections\label{sec:formfactor}}

\begin{itemize}
\item The form factors of the triangle loop $V_{1}$ \begin{eqnarray}
f_{2}^{V_{1}} & = & \frac{1}{t_{1}}\biggl\{2C_{0}m_{t}t_{1}+2m_{t}C_{11}t_{1}\biggr\},\\
f_{6}^{V_{1}} & = & \frac{1}{t_{1}}\biggl\{8C_{0}m_{t}^{2}+8m_{t}^{2}C_{11}-4m_{t}^{2}C_{21}+8C_{24}\biggr\},\\
f_{12}^{V_{1}} & = & \frac{1}{t_{1}}\biggl\{2C_{0}\left(2m_{t}^{2}+t_{1}\right)+2\left(4m_{t}^{2}+t_{1}\right)C_{11}+2t_{1}C_{12}\nonumber \\
 &  & \quad+2m_{t}^{2}C_{21}+2t_{1}C_{23}+4C_{24}\biggr\},\\
f_{16}^{V_{1}} & = & \frac{1}{t_{1}}\biggl\{4m_{t}C_{11}+4m_{t}C_{21}\biggr\},\end{eqnarray}
where the arguments of the scalar function and tensor coefficients
are $\left(\left(-p_{t}\right)^{2},p_{g}^{2},\left(p_{t}-p_{g}\right)^{2};0,m_{t}^{2},m_{t}^{2}\right)=\left(m_{t}^{2},0,t;0,m_{t}^{2},m_{t}^{2}\right)$. 
\item The form factors of the triangle loop $V_{2}$ \begin{eqnarray}
f_{2}^{V_{2}} & = & \frac{1}{t_{1}}\biggl\{3C_{0}m_{t}t_{1}+3m_{t}C_{11}t_{1}\biggr\},\\
f_{6}^{V_{2}} & = & \frac{1}{t_{1}}\biggl\{12C_{0}m_{t}^{2}+12C_{21}m_{t}^{2}+2\left(12m_{t}^{2}+t_{1}\right)C_{11}+4t_{1}C_{12}+8t_{1}C_{23}+24C_{24}\nonumber \\
 &  & +\epsilon\left(-8C_{0}m_{t}^{2}-16C_{11}m_{t}^{2}-8C_{21}m_{t}^{2}-4t_{1}C_{12}-4t_{1}C_{23}-16C_{24}\right)\biggr\},\\
f_{12}^{V_{2}} & = & \frac{1}{t_{1}}\biggl\{6C_{0}m_{t}^{2}+2C_{21}m_{t}^{2}+2\left(4m_{t}^{2}+t_{1}\right)C_{11}+t_{1}C_{12}\nonumber \\
 &  & +2t_{1}C_{23}+12C_{24}-8\epsilon C_{24}\biggr\},\\
f_{16}^{V_{2}} & = & \frac{1}{t_{1}}\biggl\{-4m_{t}C_{11}-4m_{t}C_{21}+\epsilon\left(4C_{0}m_{t}+8C_{11}m_{t}+4C_{21}m_{t}\right)\biggr\},\end{eqnarray}
where the arguments of the scalar function and tensor coefficients
are $\left(\left(-p_{2}\right)^{2},p_{g}^{2},\left(p_{t}-p_{g}\right)^{2};m_{t}^{2},0,0\right)=\left(m_{t}^{2},0,t;m_{t}^{2},0,0\right)$. 
\item The form factors of the triangle loop $V_{3}$: \begin{eqnarray}
f_{6}^{V_{3}} & = & \frac{1}{t_{1}}\biggl\{4C_{0}\left(m_{t}^{2}-m_{W}^{2}+t_{1}\right)+4\left(m_{t}^{2}-m_{W}^{2}+t_{1}\right)C_{11}+4\left(m_{t}^{2}+t_{1}\right)C_{12}\nonumber \\
 &  & +4m_{W}^{2}C_{22}+4\left(m_{t}^{2}-m_{W}^{2}+t_{1}\right)C_{23}+8C_{24}-16C_{24}\epsilon\biggr\},\\
f_{9}^{V_{3}} & = & \frac{1}{t_{1}}\biggl\{-8m_{t}C_{12}+8\epsilon m_{t}C_{23}-8m_{t}C_{23}\biggr\},\\
f_{12}^{V_{3}} & = & \frac{1}{t_{1}}\biggl\{2C_{0}\left(m_{t}^{2}-m_{W}^{2}+t_{1}\right)+2\left(m_{t}^{2}-m_{W}^{2}+t_{1}\right)C_{11}+2\left(m_{t}^{2}+t_{1}\right)C_{12}\nonumber \\
 &  & +2m_{W}^{2}C_{22}+2\left(m_{t}^{2}-m_{W}^{2}+t_{1}\right)C_{23}+4C_{24}-8C_{24}\epsilon\biggr\},\\
f_{13}^{V_{3}} & = & \frac{1}{t_{1}}\biggl\{-4m_{t}C_{12}-4m_{t}C_{23}\biggr\},\end{eqnarray}
where the arguments of the scalar function and tensor coefficients
are $\left(p_{b}^{2},\left(-p_{W}\right)^{2},\left(p_{W}-p_{b}\right)^{2};0,0,m_{t}^{2}\right)=\left(0,m_{W}^{2},t;0,0,m_{t}^{2}\right)$. 
\item The form factors of the triangle loop $V_{1}^{\prime}$ ( $f_{i}^{V_{1}^{\prime}}\equiv F_{i}^{V_{1}^{\prime}}/s$)\begin{eqnarray}
f_{4}^{V_{1}^{\prime}} & = & (8C_{24}-16C_{24}\epsilon)/s,\\
f_{5}^{V_{1}^{\prime}} & = & 4C_{0}+4C_{11}+4C_{12}+4C_{23}+8C_{24}/s+\left(-4C_{12}-4C_{23}-16C_{24}/s\right)\epsilon,\\
f_{11}^{V_{1}^{\prime}} & = & -4C_{0}-4C_{11}-4C_{12}-4C_{23}-8C_{24}/s+\left(4C_{12}+4C_{23}+16C_{24}/s\right)\epsilon,\\
f_{12}^{V_{1}^{\prime}} & = & 2C_{0}+2C_{11}+2C_{12}+2C_{23}+4C_{24}/s+\left(-2C_{12}-2C_{23}-8C_{24}/s\right)\epsilon,\end{eqnarray}
where the arguments of the scalar function and tensor coefficients
are $\left(p_{b}^{2},p_{g}^{2},\left(p_{b}+p_{g}\right)^{2},0,0,0\right)=\left(0,0,s;0,0,0\right)$. 
\item The form factors of the triangle loop $V_{2}^{\prime}$:\begin{eqnarray}
f_{4}^{V_{2}^{\prime}} & = & 2C_{11}+4C_{12}+8C_{23}+24C_{24}/s+\epsilon\left(-4C_{12}-4C_{23}-16C_{24}/s\right),\\
f_{5}^{V_{2}^{\prime}} & = & 4C_{11}+2C_{12}+4C_{23}+24C_{24}/s-16\epsilon C_{24}/s,\\
f_{11}^{V_{2}^{\prime}} & = & -4C_{11}-2C_{12}-4C_{23}-24C_{24}/s+16\epsilon C_{24}/s,\\
f_{12}^{V_{2}^{\prime}} & = & 2C_{11}+C_{12}+2C_{23}+12C_{24}/s-8\epsilon C_{24}/s,\end{eqnarray}
where the arguments of the scalar function and tensor coefficients
are $\left(p_{b}^{2},p_{g}^{2},\left(p_{b}+p_{g}\right)^{2},0,0,0\right)=\left(0,0,s;0,0,0\right)$. 
\item The form factors of the triangle loop $V_{3}^{\prime}$: \begin{eqnarray}
f_{1}^{V_{3}^{\prime}} & = & \frac{1}{s}\biggl\{4C_{0}m_{t}+4m_{t}C_{11}\biggr\},\\
f_{2}^{V_{3}^{\prime}} & = & -2C_{0}m_{t}-2m_{t}C_{11},\\
f_{4}^{V_{3}^{\prime}} & = & \frac{1}{s}\biggl\{4C_{0}\left(m_{t}^{2}-m_{W}^{2}+s\right)+4\left(2m_{t}^{2}-m_{W}^{2}+s\right)C_{11}+\left(4s-4m_{t}^{2}\right)C_{12}\nonumber \\
 &  & +4m_{t}^{2}C_{21}+4m_{W}^{2}C_{22}-4\left(m_{t}^{2}+m_{W}^{2}-s\right)C_{23}+8C_{24}-16C_{24}\epsilon\biggr\},\\
f_{5}^{V_{3}^{\prime}} & = & \frac{1}{s}\biggl\{4C_{0}\left(m_{t}^{2}-m_{W}^{2}\right)+\left(8m_{t}^{2}-4m_{W}^{2}\right)C_{11}-4m_{t}^{2}C_{12}\nonumber \\
 &  & +4m_{t}^{2}C_{21}+4m_{W}^{2}C_{22}-4\left(m_{t}^{2}+m_{W}^{2}\right)C_{23}+8C_{24}-16C_{24}\epsilon\biggr\},\\
f_{8}^{V_{3}^{\prime}} & = & \frac{1}{s}\biggl\{-8m_{t}C_{11}+8m_{t}C_{12}-8m_{t}C_{21}+8m_{t}C_{23}\biggr\},\\
f_{11}^{V_{3}^{\prime}} & = & \frac{1}{s}\biggl\{-4C_{0}\left(m_{t}^{2}-m_{W}^{2}+s\right)-4\left(2m_{t}^{2}-m_{W}^{2}+s\right)C_{11}+4\left(m_{t}^{2}-s\right)C_{12}\nonumber \\
 &  & -4m_{t}^{2}C_{21}-4m_{W}^{2}C_{22}+4\left(m_{t}^{2}+m_{W}^{2}-s\right)C_{23}-8C_{24}+16C_{24}\epsilon\biggr\},\\
f_{12}^{V_{3}^{\prime}} & = & \frac{1}{s}\biggl\{2C_{0}\left(m_{t}^{2}-m_{W}^{2}+s\right)+2\left(2m_{t}^{2}-m_{W}^{2}+s\right)C_{11}+\left(2s-2m_{t}^{2}\right)C_{12}\nonumber \\
 &  & +2m_{t}^{2}C_{21}+2m_{W}^{2}C_{22}-2\left(m_{t}^{2}+m_{W}^{2}-s\right)C_{23}+4C_{24}-8C_{24}\epsilon\biggr\},\\
f_{15}^{V_{3}^{\prime}} & = & \frac{1}{s}\biggl\{-4m_{t}C_{11}+4m_{t}C_{12}-4m_{t}C_{21}+4m_{t}C_{23}\biggr\},\end{eqnarray}
where the arguments of the scalar function and tensor coefficients
are $\left(\left(-p_{t}\right)^{2},\left(-p_{W}\right)^{2},\left(p_{t}+p_{W}\right)^{2},0,m_{t}^{2},0\right)=\left(m_{t}^{2},m_{W}^{2},s;0,m_{t}^{2},0\right)$. 
\item The form factors of the box loop $B_{1}$:\begin{eqnarray}
f_{1}^{B_{1}} & = & m_{t}\biggl\{4\left(2m_{t}^{2}+t_{1}\right)D_{21}+4\left(m_{t}^{2}+m_{W}^{2}+t_{1}\right)D_{23}+4t_{1}D_{24}\nonumber \\
 &  & -4\left(3m_{t}^{2}+m_{W}^{2}+2t_{1}\right)D_{25}-4t_{1}D_{26}+16D_{27}+4D_{31}m_{t}^{2}-4m_{W}^{2}D_{33}\nonumber \\
 &  & +4t_{1}D_{34}-4\left(2m_{t}^{2}+m_{W}^{2}-s\right)D_{35}+4\left(m_{t}^{2}+2m_{W}^{2}-s\right)D_{37}\nonumber \\
 &  & +4\left(s+t_{1}\right)D_{39}-4\left(s+2t_{1}\right)D_{310}+16D_{311}-16D_{313}\biggr\},\label{eq:fb1_1}\\
f_{2}^{B_{1}} & = & m_{t}\biggl\{-2sD_{11}+2sD_{13}-2\left(2m_{t}^{2}+t_{1}\right)D_{21}-2\left(m_{t}^{2}+m_{W}^{2}+t_{1}\right)D_{23}\nonumber \\
 &  & -2t_{1}D_{24}+2\left(3m_{t}^{2}+m_{W}^{2}+2t_{1}\right)D_{25}+2t_{1}D_{26}-8D_{27}-2D_{31}m_{t}^{2}\nonumber \\
 &  & +2m_{W}^{2}D_{33}-2t_{1}D_{34}+2\left(2m_{t}^{2}+m_{W}^{2}-s\right)D_{35}-2\left(m_{t}^{2}+2m_{W}^{2}-s\right)D_{37}\nonumber \\
 &  & -2\left(s+t_{1}\right)D_{39}+2\left(s+2t_{1}\right)D_{310}-12D_{311}+12D_{313}\nonumber \\
 &  & +\epsilon\biggl[-2D_{31}m_{t}^{2}-2\left(2m_{t}^{2}+t_{1}\right)D_{21}-2\left(m_{t}^{2}+m_{W}^{2}+t_{1}\right)D_{23}-2t_{1}D_{24}\nonumber \\
 &  & +2\left(3m_{t}^{2}+m_{W}^{2}+2t_{1}\right)D_{25}+2t_{1}D_{26}-4D_{27}+2m_{W}^{2}D_{33}-2t_{1}D_{34}\nonumber \\
 &  & +2\left(2m_{t}^{2}+m_{W}^{2}-s\right)D_{35}m_{t}-2\left(m_{t}^{2}+2m_{W}^{2}-s\right)D_{37}m_{t}-2\left(s+t_{1}\right)D_{39}m_{t}\nonumber \\
 &  & +2\left(s+2t_{1}\right)D_{310}m_{t}-8D_{311}m_{t}+8D_{313}m_{t}\biggr]+\epsilon^{2}\left(4D_{311}-4D_{313}\right)\biggr\}\label{eq:fb1_2}\\
f_{4}^{B_{1}} & = & -8m_{t}^{2}D_{11}-4t_{1}D_{12}+4\left(2m_{t}^{2}+t_{1}\right)D_{13}-4m_{t}^{2}D_{21}-4t_{1}D_{24}+4m_{t}^{2}D_{25}\nonumber \\
 &  & +4t_{1}D_{26}-8D_{27}-8D_{313}+\epsilon\biggl[-4D_{35}m_{t}^{2}+4t_{1}D_{23}-4t_{1}D_{25}+8D_{27}\nonumber \\
 &  & -4m_{W}^{2}D_{33}+4\left(m_{t}^{2}+m_{W}^{2}-s\right)D_{37}+4\left(s+t_{1}\right)D_{39}-4t_{1}D_{310}\nonumber \\
 &  & -16D_{313}\biggr]+8D_{313}\epsilon^{2},\\
f_{5}^{B_{1}} & = & -4t_{1}D_{12}+4t_{1}D_{13}-4m_{t}^{2}D_{21}-4m_{W}^{2}D_{23}-4t_{1}D_{24}+4\left(m_{t}^{2}+m_{W}^{2}\right)D_{25}\nonumber \\
 &  & +4t_{1}D_{26}-8D_{27}-8D_{311}+\epsilon\biggl[-8D_{21}m_{t}^{2}-4D_{31}m_{t}^{2}+4t_{1}D_{22}+\left(8m_{t}^{2}-4t_{1}\right)D_{24}\nonumber \\
 &  & +4\left(m_{t}^{2}+m_{W}^{2}\right)D_{25}-4\left(m_{t}^{2}+m_{W}^{2}\right)D_{26}+8D_{27}+4\left(m_{t}^{2}-t_{1}\right)D_{34}\nonumber \\
 &  & +4\left(m_{t}^{2}+m_{W}^{2}-s\right)D_{35}+4t_{1}D_{36}-4m_{W}^{2}D_{37}-4\left(s+t_{1}\right)D_{38}+4m_{W}^{2}D_{39}\nonumber \\
 &  & -4\left(m_{t}^{2}+m_{W}^{2}-2s-t_{1}\right)D_{310}-16D_{311}+16D_{312}\biggr]\nonumber \\
 &  & +8D_{312}\left(8D_{311}-8D_{312}\right)\epsilon^{2}\\
f_{6}^{B_{1}} & = & 4D_{0}\left(m_{t}^{2}-m_{W}^{2}+s+t_{1}\right)+\left(12m_{t}^{2}+8\left(-m_{W}^{2}+s+t_{1}\right)\right)D_{11}+\left(4t_{1}-4s\right)D_{12}\nonumber \\
 &  & +\left(-8m_{t}^{2}+4m_{W}^{2}-8t_{1}\right)D_{13}+4\left(3m_{t}^{2}-m_{W}^{2}+s+t_{1}\right)D_{21}+4\left(m_{t}^{2}+m_{W}^{2}+t_{1}\right)D_{23}\nonumber \\
 &  & +\left(8t_{1}-4s\right)D_{24}-8\left(2m_{t}^{2}+t_{1}\right)D_{25}-8t_{1}D_{26}+16D_{27}+4m_{t}^{2}D_{31}-4m_{W}^{2}D_{33}\nonumber \\
 &  & +4t_{1}D_{34}-4\left(2m_{t}^{2}+m_{W}^{2}-s\right)D_{35}+4\left(m_{t}^{2}+2m_{W}^{2}-s\right)D_{37}+4\left(s+t_{1}\right)D_{39}\nonumber \\
 &  & -4\left(s+2t_{1}\right)D_{310}+16D_{311}-16D_{313}\nonumber \\
 &  & +\left(4D_{21}m_{t}^{2}+4D_{23}m_{t}^{2}-4\left(2m_{t}^{2}+s\right)D_{25}+4sD_{26}-8D_{27}\right)\epsilon+8D_{27}\epsilon^{2}\label{eq:fb1_6}\\
f_{7}^{B_{1}} & = & 8m_{t}\biggl\{ D_{23}-2D_{25}+D_{26}+D_{39}-D_{310}+\epsilon\left(D_{310}-D_{39}\right)\biggr\}\label{eq:fb1_7}\\
f_{8}^{B_{1}} & = & 8m_{t}\biggl\{ D_{25}-D_{26}-D_{35}+D_{37}-D_{39}+D_{310}+\epsilon\left(D_{35}-D_{37}+D_{39}-D_{310}\right)\biggr\},\label{eq:fb1_8}\\
f_{9}^{B_{1}} & = & 8m_{t}\biggl\{-D_{11}+D_{12}-2D_{21}-D_{23}+3D_{25}-D_{34}-D_{39}+2D_{310}\nonumber \\
 &  & +\epsilon\left(D_{24}-D_{26}+D_{34}+D_{39}-2D_{310}\right)\biggr\},\label{eq:fb1_9}\\
f_{10}^{B_{1}} & = & 8m_{t}\biggl\{-D_{12}+D_{13}-D_{31}+D_{34}+2D_{35}-D_{37}+D_{39}-2D_{310}\nonumber \\
 &  & +\epsilon\left(D_{21}-D_{24}-D_{25}+D_{26}+D_{31}-D_{34}-2D_{35}+D_{37}-D_{39}+2D_{310}\right)\biggr\},\label{eq:fb1_10}\\
f_{11}^{B_{1}} & = & 4D_{21}m_{t}^{2}-8D_{24}m_{t}^{2}-4t_{1}D_{22}-4\left(m_{t}^{2}+t_{1}\right)D_{23}+4\left(m_{t}^{2}-m_{W}^{2}+s+t_{1}\right)D_{25}\nonumber \\
 &  & +4\left(m_{t}^{2}+m_{W}^{2}-s+t_{1}\right)D_{26}-16\epsilon D_{27}+4m_{W}^{2}D_{33}-4D_{34}m_{t}^{2}+4D_{35}m_{t}^{2}\nonumber \\
 &  & -4t_{1}D_{36}-4\left(m_{t}^{2}+m_{W}^{2}-s\right)D_{37}+4\left(s+t_{1}\right)D_{38}-4\left(m_{W}^{2}+s+t_{1}\right)D_{39}\nonumber \\
 &  & +4\left(m_{t}^{2}+m_{W}^{2}-s+t_{1}\right)D_{310}-16D_{312}+16D_{313}\label{eq:fb1_11}\\
f_{12}^{B_{1}} & = & 2D_{0}\left(m_{t}^{2}-m_{W}^{2}+s+t_{1}\right)+2\left(2m_{t}^{2}-m_{W}^{2}+s+t_{1}\right)D_{11}-2m_{t}^{2}D_{13}-2m_{t}^{2}D_{21}\nonumber \\
 &  & +2t_{1}D_{22}+2\left(m_{t}^{2}+t_{1}\right)D_{23}+4m_{t}^{2}D_{24}-2\left(m_{t}^{2}-m_{W}^{2}+s+t_{1}\right)D_{25}\nonumber \\
 &  & -2\left(m_{t}^{2}+m_{W}^{2}-s+t_{1}\right)D_{26}+4D_{27}-2m_{W}^{2}D_{33}+2m_{t}^{2}D_{34}-2m_{t}^{2}D_{35}\nonumber \\
 &  & +2t_{1}D_{36}+2\left(m_{t}^{2}+m_{W}^{2}-s\right)D_{37}-2\left(s+t_{1}\right)D_{38}+2\left(m_{W}^{2}+s+t_{1}\right)D_{39}\nonumber \\
 &  & -2\left(m_{t}^{2}+m_{W}^{2}-s+t_{1}\right)D_{310}+12D_{312}-12D_{313}+\left(4D_{27}-4D_{312}+4D_{313}\right)\epsilon^{2}\nonumber \\
 &  & +\epsilon\biggl[-2D_{21}m_{t}^{2}+4D_{24}m_{t}^{2}+2D_{34}m_{t}^{2}-2D_{35}m_{t}^{2}+2t_{1}D_{22}+2\left(m_{t}^{2}+t_{1}\right)D_{23}\nonumber \\
 &  & -2\left(m_{t}^{2}-m_{W}^{2}+s+t_{1}\right)D_{25}-2\left(m_{t}^{2}+m_{W}^{2}-s+t_{1}\right)D_{26}+8D_{27}-2m_{W}^{2}D_{33}\nonumber \\
 &  & +2t_{1}D_{36}+2\left(m_{t}^{2}+m_{W}^{2}-s\right)D_{37}-2\left(s+t_{1}\right)D_{38}+2\left(m_{W}^{2}+s+t_{1}\right)D_{39}\nonumber \\
 &  & -2\left(m_{t}^{2}+m_{W}^{2}-s+t_{1}\right)D_{310}+8D_{312}-8D_{313}\biggr]\label{eq:fb1_12}\\
f_{13}^{B_{1}} & = & -4m_{t}D_{11}+4m_{t}D_{12}+4m_{t}D_{24}-4m_{t}D_{26}+\epsilon\left(4m_{t}D_{24}-4m_{t}D_{26}\right)\label{eq:fb1-13}\\
f_{14}^{B_{1}} & = & 4m_{t}D_{11}-4m_{t}D_{13}-4m_{t}D_{25}+\epsilon\left(4m_{t}D_{23}-4m_{t}D_{25}\right)+4m_{t}D_{26}\label{eq:fb1_14}\\
f_{15}^{B_{1}} & = & -4m_{t}D_{12}+4m_{t}D_{13}+4m_{t}D_{21}-4m_{t}D_{24}-4m_{t}D_{25}+4m_{t}D_{26}\nonumber \\
 &  & +\epsilon\left(4m_{t}D_{21}-4m_{t}D_{24}-4m_{t}D_{25}+4m_{t}D_{26}\right)\label{eq:fb1_15}\\
f_{16}^{B_{1}} & = & 4m_{t}D_{12}-4m_{t}D_{13}-4m_{t}D_{21}+4m_{t}D_{24}+4m_{t}D_{25}\nonumber \\
 &  & +\epsilon\left(-4m_{t}D_{21}-4m_{t}D_{23}+8m_{t}D_{25}\right)-4m_{t}D_{26}\label{eq:fb1_16}\\
f_{17}^{B_{1}} & = & -8D_{23}+8D_{26}+8D_{38}-8D_{39}+\epsilon\left(8D_{23}-8D_{26}-8D_{38}+8D_{39}\right)\label{eq:fb1_17}\\
f_{18}^{B_{1}} & = & -8D_{25}+8D_{26}-8D_{37}-8D_{38}+8D_{39}\nonumber \\
 &  & +\epsilon\left(-8D_{25}+8D_{26}+8D_{37}+8D_{38}-8D_{39}-8D_{310}\right)+8D_{310}\label{eq:fb1_18}\\
f_{19}^{B_{1}} & = & 16D_{12}-16D_{13}+8D_{22}+8D_{23}+16D_{24}-8D_{25}-24D_{26}+8D_{36}-8D_{38}+8D_{39}\nonumber \\
 &  & -8D_{310}+\epsilon\left(-8D_{22}-8D_{23}+8D_{25}+8D_{26}-8D_{36}+8D_{38}-8D_{39}+8D_{310}\right)\label{eq:fb1_19}\\
f_{20}^{B_{1}} & = & -8D_{22}+8D_{24}-8D_{25}+8D_{26}+8D_{34}-8D_{35}-8D_{36}+8D_{37}+8D_{38}-8D_{39}\nonumber \\
 &  & +\epsilon\biggl[8D_{22}-8D_{24}+8D_{25}-8D_{26}-8D_{34}+8D_{35}+8D_{36}\nonumber \\
 &  & -8D_{37}-8D_{38}+8D_{39}\biggr],\end{eqnarray}
where the arguments of the scalar function and tensor coefficients
are $\left(\left(-p_{t}\right)^{2},p_{g}^{2},\left(-p_{W}\right)^{2},p_{b}^{2};t,u;0,m_{t}^{2},m_{t}^{2},0\right)=\left(m_{t}^{2},0,m_{W}^{2},0;t,u;0,m_{t}^{2},m_{t}^{2},0\right)$. 
\item The form factors of the box loop $B_{2}$\begin{eqnarray}
f_{1}^{B_{2}} & = & -4D_{21}m_{t}^{3}-4\left(m_{t}^{2}-m_{W}^{2}+t_{1}\right)D_{11}m_{t}+4\left(m_{t}^{2}-m_{W}^{2}-s+t_{1}\right)D_{12}m_{t}+4sD_{13}m_{t}\nonumber \\
 &  & -4m_{W}^{2}D_{22}m_{t}+4\left(m_{t}^{2}+m_{W}^{2}-s\right)D_{24}m_{t}-4t_{1}D_{25}m_{t}+4\left(s+t_{1}\right)D_{26}m_{t}-8D_{27}m_{t}\nonumber \\
 &  & -8D_{311}m_{t}+8D_{312}m_{t}+\epsilon^{2}\left(8m_{t}D_{311}-8m_{t}D_{312}\right)+\epsilon\biggl[-4D_{31}m_{t}^{3}\nonumber \\
 &  & -4\left(m_{t}^{2}-m_{W}^{2}+s\right)D_{21}m_{t}+4\left(-m_{t}^{2}+m_{W}^{2}+s\right)D_{22}m_{t}+8\left(m_{t}^{2}-m_{W}^{2}\right)D_{24}m_{t}\nonumber \\
 &  & +4sD_{25}m_{t}-4sD_{26}m_{t}+8D_{27}m_{t}+4m_{W}^{2}D_{32}m_{t}+4\left(2m_{t}^{2}+m_{W}^{2}-s\right)D_{34}m_{t}\nonumber \\
 &  & -4t_{1}D_{35}m_{t}-4\left(m_{t}^{2}+2m_{W}^{2}-s\right)D_{36}m_{t}-4\left(s+t_{1}\right)D_{38}m_{t}+4\left(s+2t_{1}\right)D_{310}m_{t}\nonumber \\
 &  & -16D_{311}m_{t}+16D_{312}m_{t}\biggr],\label{eq:fb2_1}\\
f_{2}^{B_{2}} & = & 2D_{31}m_{t}^{3}+2\left(m_{t}^{2}-m_{W}^{2}+t_{1}\right)D_{11}m_{t}+2\left(-m_{t}^{2}+m_{W}^{2}+s-t_{1}\right)D_{12}m_{t}-2sD_{13}m_{t}\nonumber \\
 &  & +2\left(2m_{t}^{2}-m_{W}^{2}+s\right)D_{21}m_{t}+2\left(m_{t}^{2}-s\right)D_{22}m_{t}+2\left(-3m_{t}^{2}+m_{W}^{2}+s\right)D_{24}m_{t}\nonumber \\
 &  & +2\left(t_{1}-s\right)D_{25}m_{t}-2t_{1}D_{26}m_{t}+4D_{27}m_{t}-2m_{W}^{2}D_{32}m_{t}-2\left(2m_{t}^{2}+m_{W}^{2}-s\right)D_{34}m_{t}\nonumber \\
 &  & +2t_{1}D_{35}m_{t}+2\left(m_{t}^{2}+2m_{W}^{2}-s\right)D_{36}m_{t}+2\left(s+t_{1}\right)D_{38}m_{t}-2\left(s+2t_{1}\right)D_{310}m_{t}\nonumber \\
 &  & +12D_{311}m_{t}-12D_{312}m_{t}+\epsilon\biggl[2D_{31}m_{t}^{3}+2\left(m_{t}^{2}-m_{W}^{2}+s\right)D_{21}m_{t}\nonumber \\
 &  & +2\left(m_{t}^{2}-m_{W}^{2}-s\right)D_{22}m_{t}-2sD_{25}m_{t}+2sD_{26}m_{t}-4D_{27}m_{t}-2m_{W}^{2}D_{32}m_{t}\nonumber \\
 &  & -2\left(2m_{t}^{2}+m_{W}^{2}-s\right)D_{34}m_{t}+2t_{1}D_{35}m_{t}+2\left(m_{t}^{2}+2m_{W}^{2}-s\right)D_{36}m_{t}\nonumber \\
 &  & +2\left(s+t_{1}\right)D_{38}m_{t}-2\left(s+2t_{1}\right)D_{310}m_{t}+8D_{311}m_{t}-8D_{312}m_{t}\nonumber \\
 &  & +\left(4m_{t}m_{W}^{2}-4m_{t}^{3}\right)D_{24}\biggr]+\epsilon^{2}\left(4m_{t}D_{312}-4m_{t}D_{311}\right),\label{eq:fb2_2}\\
f_{4}^{B_{2}} & = & 4D_{0}\left(m_{t}^{2}-m_{W}^{2}+s+t_{1}\right)+4\left(2m_{t}^{2}-m_{W}^{2}+s+t_{1}\right)D_{11}+\left(8s-4m_{W}^{2}\right)D_{12}\nonumber \\
 &  & +\left(4t_{1}-4s\right)D_{13}+4m_{t}^{2}D_{21}+4\left(-m_{t}^{2}+m_{W}^{2}+s\right)D_{22}+4\left(m_{t}^{2}-2m_{W}^{2}+2s\right)D_{24}\nonumber \\
 &  & +4t_{1}D_{25}-8sD_{26}+16D_{27}+4m_{W}^{2}D_{32}+4m_{t}^{2}D_{34}-4\left(m_{t}^{2}+m_{W}^{2}-s\right)D_{36}\nonumber \\
 &  & -4\left(s+t_{1}\right)D_{38}+4t_{1}D_{310}+16D_{312}\nonumber \\
 &  & +\left(-4D_{21}m_{t}^{2}-4D_{22}m_{t}^{2}+8D_{24}m_{t}^{2}-4t_{1}D_{25}+4t_{1}D_{26}-8D_{27}\right)\epsilon+8D_{27}\epsilon^{2},\label{eq:fb2_4}\\
f_{5}^{B_{2}} & = & 4D_{0}\left(m_{t}^{2}-m_{W}^{2}+s+t_{1}\right)+\left(12m_{t}^{2}+8\left(-m_{W}^{2}+s+t_{1}\right)\right)D_{11}\nonumber \\
 &  & +4\left(-2m_{t}^{2}+m_{W}^{2}+s-t_{1}\right)D_{12}-8sD_{13}+4\left(2m_{t}^{2}-m_{W}^{2}+s+t_{1}\right)D_{21}+4m_{W}^{2}D_{22}\nonumber \\
 &  & +4sD_{23}-4\left(2m_{t}^{2}-2s+t_{1}\right)D_{24}-4\left(m_{t}^{2}-m_{W}^{2}+2s\right)D_{25}+4\left(m_{t}^{2}-m_{W}^{2}-2s\right)D_{26}\nonumber \\
 &  & +16D_{27}+4m_{t}^{2}D_{31}-4\left(m_{t}^{2}+m_{W}^{2}-s\right)D_{34}+\left(4t_{1}-4m_{t}^{2}\right)D_{35}+4m_{W}^{2}D_{36}-4t_{1}D_{37}\nonumber \\
 &  & -4m_{W}^{2}D_{38}+4\left(s+t_{1}\right)D_{39}+4\left(m_{t}^{2}+m_{W}^{2}-2s-t_{1}\right)D_{310}+16D_{311}-16D_{313}\nonumber \\
 &  & +\left(-4D_{21}m_{t}^{2}-4m_{W}^{2}D_{22}+4\left(m_{t}^{2}+m_{W}^{2}\right)D_{24}-4t_{1}D_{25}+4t_{1}D_{26}-8D_{27}\right)\epsilon\nonumber \\
 &  & +8D_{27}\epsilon^{2},\label{eq:fb2_5}\\
f_{6}^{B_{2}} & = & -4sD_{12}+4sD_{13}-4sD_{24}+4sD_{26}-8D_{27}-8D_{311}+8D_{312}+\left(8D_{311}-8D_{312}\right)\epsilon^{2}\nonumber \\
 &  & +\epsilon\biggl[-4D_{31}m_{t}^{2}+4sD_{22}-4sD_{24}+8D_{27}+4m_{W}^{2}D_{32}+4\left(2m_{t}^{2}+m_{W}^{2}-s\right)D_{34}\nonumber \\
 &  & -4t_{1}D_{35}-4\left(m_{t}^{2}+2m_{W}^{2}-s\right)D_{36}-4\left(s+t_{1}\right)D_{38}+4\left(s+2t_{1}\right)D_{310}\nonumber \\
 &  & -16D_{311}+16D_{312}\biggr],\label{eq:fb2_6}\\
f_{7}^{B_{2}} & = & -8m_{t}D_{12}+8m_{t}D_{13}-8m_{t}D_{24}-8m_{t}D_{25}+16m_{t}D_{26}+8m_{t}D_{38}-8m_{t}D_{310}\nonumber \\
 &  & +\epsilon\left(8m_{t}D_{22}-8m_{t}D_{24}+8m_{t}D_{25}-8m_{t}D_{26}-8m_{t}D_{38}+8m_{t}D_{310}\right),\label{eq:fb2_7}\\
f_{8}^{B_{2}} & = & -8m_{t}D_{11}+16m_{t}D_{12}-8m_{t}D_{13}-8m_{t}D_{21}+8m_{t}D_{22}+8m_{t}D_{24}+8m_{t}D_{25}\nonumber \\
 &  & -16m_{t}D_{26}-8m_{t}D_{34}+8m_{t}D_{36}-8m_{t}D_{38}+8m_{t}D_{310}+\epsilon\biggl[8m_{t}D_{21}-8m_{t}D_{22}\nonumber \\
 &  & -8m_{t}D_{25}+8m_{t}D_{26}+8m_{t}D_{34}-8m_{t}D_{36}+8m_{t}D_{38}-8m_{t}D_{310}\biggr],\label{eq:fb2_8}\\
f_{9}^{B_{2}} & = & -8m_{t}D_{21}+8m_{t}D_{24}+8m_{t}D_{25}-8m_{t}D_{26}-8m_{t}D_{35}-8m_{t}D_{38}+16m_{t}D_{310}\nonumber \\
 &  & +\epsilon\left(-8m_{t}D_{21}-8m_{t}D_{22}+16m_{t}D_{24}+8m_{t}D_{35}+8m_{t}D_{38}-16m_{t}D_{310}\right),\label{eq:fb2_9}\\
f_{10}^{B_{2}} & = & -8m_{t}D_{22}+8m_{t}D_{24}-8m_{t}D_{25}+8m_{t}D_{26}-8m_{t}D_{31}+16m_{t}D_{34}+8m_{t}D_{35}\nonumber \\
 &  & -8m_{t}D_{36}+8m_{t}D_{38}-16m_{t}D_{310}+\epsilon\biggl[8m_{t}D_{21}+8m_{t}D_{22}-16m_{t}D_{24}\nonumber \\
 &  & +8m_{t}D_{31}-16m_{t}D_{34}-8m_{t}D_{35}+8m_{t}D_{36}-8m_{t}D_{38}+16m_{t}D_{310}\biggr],\label{eq:fb2_10}\\
f_{11}^{B_{2}} & = & \left(-8D_{27}+8D_{312}-8D_{313}\right)\epsilon^{2}\nonumber \\
 &  & +\epsilon\biggl[4D_{21}m_{t}^{2}-8D_{24}m_{t}^{2}-4D_{34}m_{t}^{2}+4D_{35}m_{t}^{2}+4\left(m_{t}^{2}-s\right)D_{22}-4sD_{23}\nonumber \\
 &  & +4\left(m_{t}^{2}-m_{W}^{2}+s+t_{1}\right)D_{25}-4\left(m_{t}^{2}-m_{W}^{2}-s+t_{1}\right)D_{26}-4m_{W}^{2}D_{32}\nonumber \\
 &  & +4\left(m_{t}^{2}+m_{W}^{2}-s\right)D_{36}+4t_{1}D_{37}+4\left(m_{W}^{2}+s+t_{1}\right)D_{38}-4\left(s+t_{1}\right)D_{39}\nonumber \\
 &  & -4\left(m_{t}^{2}+m_{W}^{2}-s+t_{1}\right)D_{310}-16D_{312}+16D_{313}\biggr]\nonumber \\
 &  & -4D_{0}\left(m_{t}^{2}-m_{W}^{2}+s+t_{1}\right)-4\left(2m_{t}^{2}-m_{W}^{2}+s+t_{1}\right)D_{11}+4m_{t}^{2}D_{12}\nonumber \\
 &  & -8D_{27}-8D_{312}+8D_{313}\label{eq:fb2_11}\\
f_{12}^{B_{2}} & = & \epsilon\biggl[-2D_{21}m_{t}^{2}+4D_{24}m_{t}^{2}+2D_{34}m_{t}^{2}-2D_{35}m_{t}^{2}+\left(2s-2m_{t}^{2}\right)D_{22}+2sD_{23}\nonumber \\
 &  & -2\left(m_{t}^{2}-m_{W}^{2}+s+t_{1}\right)D_{25}+2\left(m_{t}^{2}-m_{W}^{2}-s+t_{1}\right)D_{26}+8D_{27}+2m_{W}^{2}D_{32}\nonumber \\
 &  & -2\left(m_{t}^{2}+m_{W}^{2}-s\right)D_{36}-2t_{1}D_{37}-2\left(m_{W}^{2}+s+t_{1}\right)D_{38}+2\left(s+t_{1}\right)D_{39}\nonumber \\
 &  & +2\left(m_{t}^{2}+m_{W}^{2}-s+t_{1}\right)D_{310}+8D_{312}-8D_{313}\biggr]\nonumber \\
 &  & +2D_{0}\left(m_{t}^{2}-m_{W}^{2}+s+t_{1}\right)+2\left(2m_{t}^{2}-m_{W}^{2}+s+t_{1}\right)D_{11}-2m_{t}^{2}D_{12}-2m_{t}^{2}D_{21}\nonumber \\
 &  & +\left(2s-2m_{t}^{2}\right)D_{22}+2sD_{23}+4m_{t}^{2}D_{24}-2\left(m_{t}^{2}-m_{W}^{2}+s+t_{1}\right)D_{25}\nonumber \\
 &  & +2\left(m_{t}^{2}-m_{W}^{2}-s+t_{1}\right)D_{26}+4D_{27}+2m_{W}^{2}D_{32}+2m_{t}^{2}D_{34}-2m_{t}^{2}D_{35}\nonumber \\
 &  & -2\left(m_{t}^{2}+m_{W}^{2}-s\right)D_{36}-2t_{1}D_{37}-2\left(m_{W}^{2}+s+t_{1}\right)D_{38}+2\left(s+t_{1}\right)D_{39}\nonumber \\
 &  & +2\left(m_{t}^{2}+m_{W}^{2}-s+t_{1}\right)D_{310}+12D_{312}-12D_{313}\nonumber \\
 &  & +\left(4D_{27}-4D_{312}+4D_{313}\right)\epsilon^{2}\\
f_{13}^{B_{2}} & = & -4m_{t}D_{11}+4m_{t}D_{12}+4m_{t}D_{25}-4m_{t}D_{26}+\epsilon\left(4m_{t}D_{25}-4m_{t}D_{26}\right)\label{eq:fb2_13}\\
f_{14}^{B_{2}} & = & 4m_{t}D_{11}-8m_{t}D_{12}+4m_{t}D_{13}-4m_{t}D_{24}+\epsilon\left(4m_{t}D_{22}-4m_{t}D_{24}\right)+4m_{t}D_{26}\label{eq:fb2_14}\\
f_{15}^{B_{2}} & = & 4m_{t}D_{21}-4m_{t}D_{24}-4m_{t}D_{25}+4m_{t}D_{26}\nonumber \\
 &  & +\epsilon\left(4m_{t}D_{21}-4m_{t}D_{24}-4m_{t}D_{25}+4m_{t}D_{26}\right)\label{eq:fb2_15}\\
f_{16}^{B_{2}} & = & -4m_{t}D_{21}+4m_{t}D_{24}+\epsilon\left(-4m_{t}D_{21}-4m_{t}D_{22}+8m_{t}D_{24}\right)+4m_{t}D_{25}-4m_{t}D_{26}\label{eq:fb2_16}\\
f_{17}^{B_{2}} & = & 8D_{23}-8D_{26}-8D_{38}+\epsilon\left(-8D_{23}+8D_{26}+8D_{38}-8D_{39}\right)+8D_{39}\label{eq:fb2_17}\\
f_{18}^{B_{2}} & = & -16D_{12}+16D_{13}-8D_{22}-8D_{23}-16D_{24}+8D_{25}+24D_{26}-8D_{36}+8D_{38}-8D_{39}\nonumber \\
 &  & +\epsilon\left(8D_{22}+8D_{23}-8D_{25}-8D_{26}+8D_{36}-8D_{38}+8D_{39}-8D_{310}\right)+8D_{310}\label{eq:fb2_18}\\
f_{19}^{B_{2}} & = & 8D_{25}-8D_{26}+8D_{37}+8D_{38}-8D_{39}-8D_{310}\nonumber \\
 &  & +\epsilon\left(8D_{25}-8D_{26}-8D_{37}-8D_{38}+8D_{39}+8D_{310}\right)\label{eq:fb2_19}\\
f_{20}^{B_{2}} & = & 8D_{22}-8D_{24}+8D_{25}-8D_{26}-8D_{34}+8D_{35}+8D_{36}-8D_{37}-8D_{38}+8D_{39}\nonumber \\
 &  & +\epsilon\biggl[-8D_{22}+8D_{24}-8D_{25}+8D_{26}+8D_{34}-8D_{35}-8D_{36}\nonumber \\
 &  & +8D_{37}+8D_{38}-8D_{39}\biggr],\label{eq:fb2_20}\end{eqnarray}
where the arguments of the scalar function and tensor coefficients
are $\left(\left(-p_{t}\right)^{2},\left(-p_{W}\right)^{2},p_{g}^{2},p_{b}^{2};s,u;0,m_{t}^{2},0,0\right)=\left(m_{t}^{2},m_{W}^{2},0,0;s,u;0,m_{t}^{2},0,0\right)$. 
\item The form factors of the box loop $B_{3}$:\begin{eqnarray}
f_{1}^{B_{3}} & = & -4D_{21}m_{t}^{3}-2D_{31}m_{t}^{3}+4t_{1}D_{11}m_{t}-4t_{1}D_{12}m_{t}-2\left(m_{t}^{2}+m_{W}^{2}-s\right)D_{22}m_{t}\nonumber \\
 &  & +2\left(3m_{t}^{2}+m_{W}^{2}-s\right)D_{24}m_{t}+2\left(m_{t}^{2}-m_{W}^{2}+s+t_{1}\right)D_{25}m_{t}\nonumber \\
 &  & -2\left(m_{t}^{2}-m_{W}^{2}+s+t_{1}\right)D_{26}m_{t}-12D_{27}m_{t}+2m_{W}^{2}D_{32}m_{t}\nonumber \\
 &  & +2\left(2m_{t}^{2}+m_{W}^{2}-s\right)D_{34}m_{t}+2\left(m_{t}^{2}-m_{W}^{2}+s+t_{1}\right)D_{35}m_{t}\nonumber \\
 &  & -2\left(m_{t}^{2}+2m_{W}^{2}-s\right)D_{36}m_{t}+2\left(m_{t}^{2}-m_{W}^{2}+t_{1}\right)D_{38}m_{t}\nonumber \\
 &  & -2\left(2m_{t}^{2}-2m_{W}^{2}+s+2t_{1}\right)D_{310}m_{t}-4D_{311}m_{t}+4D_{312}m_{t}\nonumber \\
 &  & +\epsilon\left(4m_{t}D_{312}-4m_{t}D_{311}\right)\label{eq:fb3_1}\\
f_{2}^{B_{3}} & = & 2m_{t}D_{22}m_{W}^{2}+m_{t}\left(s-2t_{1}\right)D_{11}-m_{t}\left(s-2t_{1}\right)D_{12}+m_{t}\left(m_{t}^{2}+m_{W}^{2}-s\right)D_{21}\nonumber \\
 &  & +m_{t}\left(-m_{t}^{2}-3m_{W}^{2}+s\right)D_{24}-m_{t}\left(m_{t}^{2}-m_{W}^{2}+t_{1}\right)D_{25}\nonumber \\
 &  & +m_{t}\left(m_{t}^{2}-m_{W}^{2}+t_{1}\right)D_{26}+6m_{t}D_{27}\label{eq:fb3_2}\\
f_{4}^{B_{3}} & = & -4D_{0}t_{1}+4\left(m_{t}^{2}-t_{1}\right)D_{11}-2\left(3m_{t}^{2}-m_{W}^{2}+s\right)D_{12}+2\left(m_{t}^{2}-m_{W}^{2}+s+t_{1}\right)D_{13}\nonumber \\
 &  & +2m_{t}^{2}D_{21}+2\left(m_{t}^{2}-s\right)D_{22}-2\left(m_{t}^{2}-m_{W}^{2}+s+t_{1}\right)D_{23}-2\left(3m_{t}^{2}-m_{W}^{2}+s\right)D_{24}\nonumber \\
 &  & +2\left(m_{t}^{2}-m_{W}^{2}+s-t_{1}\right)D_{25}+2\left(m_{t}^{2}-m_{W}^{2}+2\left(s+t_{1}\right)\right)D_{26}-2m_{W}^{2}D_{32}\nonumber \\
 &  & -2m_{t}^{2}D_{34}+2m_{t}^{2}D_{35}+2\left(m_{t}^{2}+m_{W}^{2}-s\right)D_{36}-2\left(m_{t}^{2}-m_{W}^{2}+s+t_{1}\right)D_{37}\nonumber \\
 &  & -2\left(m_{t}^{2}-2m_{W}^{2}+t_{1}\right)D_{38}+2\left(m_{t}^{2}-m_{W}^{2}+t_{1}\right)D_{39}+\left(-4m_{W}^{2}+4s+2t_{1}\right)D_{310}\nonumber \\
 &  & -4D_{312}+4D_{313}+\epsilon\biggl[4D_{34}m_{t}^{2}-4D_{35}m_{t}^{2}+4sD_{22}+4sD_{23}-8sD_{26}\nonumber \\
 &  & -4D_{27}+4m_{W}^{2}D_{32}-4\left(m_{t}^{2}+m_{W}^{2}-s\right)D_{36}+4\left(m_{t}^{2}-m_{W}^{2}+s+t_{1}\right)D_{37}\nonumber \\
 &  & +4\left(m_{t}^{2}-2m_{W}^{2}+t_{1}\right)D_{38}-4\left(m_{t}^{2}-m_{W}^{2}+t_{1}\right)D_{39}+\left(8m_{W}^{2}-4\left(2s+t_{1}\right)\right)D_{310}\nonumber \\
 &  & +20D_{312}-20D_{313}\biggr]+\left(8D_{313}-8D_{312}\right)\epsilon^{2}\label{eq:fb3_4}\\
f_{5}^{B_{3}} & = & 2D_{31}m_{t}^{2}-4D_{0}t_{1}+4\left(m_{t}^{2}-2t_{1}\right)D_{11}-2\left(m_{t}^{2}+m_{W}^{2}-s-2t_{1}\right)D_{12}\nonumber \\
 &  & -2\left(m_{t}^{2}-m_{W}^{2}+s+t_{1}\right)D_{13}+\left(10m_{t}^{2}-4t_{1}\right)D_{21}+6m_{W}^{2}D_{22}\nonumber \\
 &  & +4\left(-2m_{t}^{2}-2m_{W}^{2}+s+t_{1}\right)D_{24}-4\left(2m_{t}^{2}-2m_{W}^{2}+s+2t_{1}\right)D_{25}\nonumber \\
 &  & +6\left(m_{t}^{2}-m_{W}^{2}+t_{1}\right)D_{26}+28D_{27}-2\left(m_{t}^{2}+m_{W}^{2}-s\right)D_{34}\nonumber \\
 &  & -2\left(m_{t}^{2}-m_{W}^{2}+s+t_{1}\right)D_{35}+2m_{W}^{2}D_{36}+2\left(m_{t}^{2}-m_{W}^{2}+t_{1}\right)D_{310}\nonumber \\
 &  & +\epsilon\left(-20D_{27}-12D_{311}\right)+20D_{311}\label{eq:fb3_5}\\
f_{6}^{B_{3}} & = & -2D_{0}s-2\left(m_{t}^{2}-m_{W}^{2}+2s\right)D_{11}+2\left(m_{t}^{2}-m_{W}^{2}+2s\right)D_{12}+2sD_{13}\nonumber \\
 &  & -2\left(2m_{t}^{2}-m_{W}^{2}+s\right)D_{21}+\left(2s-2m_{t}^{2}\right)D_{22}+2\left(3m_{t}^{2}-m_{W}^{2}+s\right)D_{24}\nonumber \\
 &  & +2\left(m_{t}^{2}-m_{W}^{2}+t_{1}\right)D_{25}-2\left(m_{t}^{2}-m_{W}^{2}+s+t_{1}\right)D_{26}-2m_{t}^{2}D_{31}+2m_{W}^{2}D_{32}\nonumber \\
 &  & +2\left(2m_{t}^{2}+m_{W}^{2}-s\right)D_{34}+2\left(m_{t}^{2}-m_{W}^{2}+s+t_{1}\right)D_{35}-2\left(m_{t}^{2}+2m_{W}^{2}-s\right)D_{36}\nonumber \\
 &  & +2\left(m_{t}^{2}-m_{W}^{2}+t_{1}\right)D_{38}-2\left(2m_{t}^{2}-2m_{W}^{2}+s+2t_{1}\right)D_{310}-4D_{311}+4D_{312}\nonumber \\
 &  & +\epsilon\biggl[4D_{31}m_{t}^{2}-4sD_{22}+4sD_{24}-4sD_{25}+4sD_{26}-4D_{27}-4m_{W}^{2}D_{32}\nonumber \\
 &  & -4\left(2m_{t}^{2}+m_{W}^{2}-s\right)D_{34}-4\left(m_{t}^{2}-m_{W}^{2}+s+t_{1}\right)D_{35}\nonumber \\
 &  & +4\left(m_{t}^{2}+2m_{W}^{2}-s\right)D_{36}-4\left(m_{t}^{2}-m_{W}^{2}+t_{1}\right)D_{38}\nonumber \\
 &  & +4\left(2m_{t}^{2}-2m_{W}^{2}+s+2t_{1}\right)D_{310}+20D_{311}-20D_{312}\biggr]\nonumber \\
 &  & +\left(8D_{312}-8D_{311}\right)\epsilon^{2}\\
f_{7}^{B_{3}} & = & 8m_{t}D_{12}-8m_{t}D_{13}+8m_{t}D_{24}-8m_{t}D_{25}+8m_{t}D_{37}+8m_{t}D_{38}\nonumber \\
 &  & -8m_{t}D_{39}-8m_{t}D_{310}+\epsilon\biggl(-8m_{t}D_{22}+8m_{t}D_{24}-8m_{t}D_{25}\nonumber \\
 &  & +8m_{t}D_{26}-8m_{t}D_{37}-8m_{t}D_{38}+8m_{t}D_{39}+8m_{t}D_{310}\biggr)\label{eq:fb3_7}\\
f_{8}^{B_{3}} & = & -8m_{t}D_{12}+8m_{t}D_{13}-8m_{t}D_{22}+8m_{t}D_{26}+8m_{t}D_{34}-8m_{t}D_{35}\nonumber \\
 &  & -8m_{t}D_{36}+8m_{t}D_{310}+\epsilon\biggl(8m_{t}D_{22}-8m_{t}D_{24}+8m_{t}D_{25}-8m_{t}D_{26}\nonumber \\
 &  & -8m_{t}D_{34}+8m_{t}D_{35}+8m_{t}D_{36}-8m_{t}D_{310}\biggr)\label{eq:fb3_8}\\
f_{9}^{B_{3}} & = & 8m_{t}D_{21}-8m_{t}D_{24}-8m_{t}D_{35}-8m_{t}D_{38}+16m_{t}D_{310}\nonumber \\
 &  & +\epsilon\left(8m_{t}D_{21}+8m_{t}D_{22}-16m_{t}D_{24}+8m_{t}D_{35}+8m_{t}D_{38}-16m_{t}D_{310}\right)\label{eq:fb3_9}\\
f_{10}^{B_{3}} & = & 8m_{t}D_{22}-8m_{t}D_{24}+8m_{t}D_{31}-16m_{t}D_{34}+8m_{t}D_{36}\nonumber \\
 &  & +\epsilon\left(-8m_{t}D_{21}-8m_{t}D_{22}+16m_{t}D_{24}-8m_{t}D_{31}+16m_{t}D_{34}-8m_{t}D_{36}\right)\label{eq:fb3_10}\\
f_{11}^{B_{3}} & = & -6D_{21}m_{t}^{2}-2D_{34}m_{t}^{2}+4D_{0}t_{1}+\left(4t_{1}-4m_{t}^{2}\right)D_{11}+2\left(m_{t}^{2}+m_{W}^{2}-3s\right)D_{12}\nonumber \\
 &  & +2\left(m_{t}^{2}-m_{W}^{2}+3s+t_{1}\right)D_{13}+2\left(m_{t}^{2}-2m_{W}^{2}-s\right)D_{22}+2\left(m_{t}^{2}+3m_{W}^{2}-3s\right)D_{24}\nonumber \\
 &  & +6\left(m_{t}^{2}-m_{W}^{2}+s+t_{1}\right)D_{25}+\left(-4m_{t}^{2}+4m_{W}^{2}+2s-4t_{1}\right)D_{26}-8D_{27}-2m_{W}^{2}D_{32}\nonumber \\
 &  & +2\left(m_{t}^{2}+m_{W}^{2}-s\right)D_{36}-2\left(m_{t}^{2}-m_{W}^{2}+t_{1}\right)D_{38}+2\left(m_{t}^{2}-m_{W}^{2}+s+t_{1}\right)D_{310}\nonumber \\
 &  & +\epsilon\left(12D_{27}-4D_{312}\right)-4D_{312}\label{eq:fb3_11}\\
f_{12}^{B_{3}} & = & 5D_{21}m_{t}^{2}-2D_{0}t_{1}+2\left(m_{t}^{2}-t_{1}\right)D_{11}+\left(-m_{t}^{2}-m_{W}^{2}+4s\right)D_{12}\nonumber \\
 &  & +\left(-m_{t}^{2}+m_{W}^{2}-3s-t_{1}\right)D_{13}+3m_{W}^{2}D_{22}-4\left(m_{t}^{2}+m_{W}^{2}-s\right)D_{24}\nonumber \\
 &  & -5\left(m_{t}^{2}-m_{W}^{2}+s+t_{1}\right)D_{25}+4\left(m_{t}^{2}-m_{W}^{2}+t_{1}\right)D_{26}-14\epsilon D_{27}+6D_{27}\label{eq:fb3_12}\\
f_{13}^{B_{3}} & = & 4m_{t}D_{11}-4m_{t}D_{12}+8m_{t}D_{25}-8m_{t}D_{26}\label{eq:fb3_13}\\
f_{14}^{B_{3}} & = & -4m_{t}D_{11}+8m_{t}D_{12}-4m_{t}D_{13}+4m_{t}D_{24}-4m_{t}D_{25}\nonumber \\
 &  & +\epsilon\left(-4m_{t}D_{22}+4m_{t}D_{24}-4m_{t}D_{25}+4m_{t}D_{26}\right)\label{eq:fb3_14}\\
f_{15}^{B_{3}} & = & -4m_{t}D_{11}+4m_{t}D_{12}-8m_{t}D_{21}+8m_{t}D_{24}\label{eq:fb3_15}\\
f_{16}^{B_{3}} & = & 4m_{t}D_{21}-4m_{t}D_{24}+\epsilon\left(4m_{t}D_{21}+4m_{t}D_{22}-8m_{t}D_{24}\right)\label{eq:fb3_16}\\
f_{17}^{B_{3}} & = & 8D_{23}-8D_{26}-8D_{38}+\epsilon\left(-8D_{23}+8D_{26}+8D_{38}-8D_{39}\right)+8D_{39}\label{eq:fb3_17}\\
f_{18}^{B_{3}} & = & 16D_{12}-16D_{13}+8D_{22}+16D_{24}-16D_{25}-8D_{26}+8D_{36}-8D_{310}\nonumber \\
 &  & +\epsilon\left(-8D_{22}+8D_{26}-8D_{36}+8D_{310}\right)\label{eq:fb3_18}\\
f_{19}^{B_{3}} & = & 8D_{25}-8D_{26}+8D_{38}-8D_{310}+\epsilon\left(8D_{25}-8D_{26}-8D_{38}+8D_{310}\right)\label{eq:fb3_19}\\
f_{20}^{B_{3}} & = & -8D_{22}+8D_{24}+8D_{34}-8D_{36}+\epsilon\left(8D_{22}-8D_{24}-8D_{34}+8D_{36}\right),\label{eq:fb3_20}\end{eqnarray}
where the arguments of the scalar function and tensor coefficients
are $\left(\left(-p_{t}\right)^{2},\left(-p_{W}\right)^{2},p_{b}^{2},p_{g}^{2};s,t;0,m_{t}^{2},0,0\right)=\left(m_{t}^{2},m_{W}^{2},0,0;s,t;0,m_{t}^{2},0,0\right)$. 
\item Bubble correction $S_{1}$:\begin{eqnarray}
f_{2}^{S_{3}} & = & \frac{1}{t_{1}}\biggl\{4B_{0}m_{t}+2B_{1}m_{t}+\epsilon\left(-2B_{0}m_{t}-2B_{1}m_{t}\right)\biggr\},\label{eq:fs1_2}\\
f_{6}^{S_{3}} & = & \frac{1}{t_{1}^{2}}\biggl\{16B_{0}m_{t}^{2}+4B_{1}\left(2m_{t}^{2}+t_{1}\right)+\epsilon\left(-8B_{0}m_{t}^{2}-4B_{1}\left(2m_{t}^{2}+t_{1}\right)\right)\biggr\},\label{eq:fs1_6}\\
f_{12}^{S_{3}} & = & \frac{1}{t_{1}^{2}}\biggl\{8B_{0}m_{t}^{2}+2B_{1}\left(2m_{t}^{2}+t_{1}\right)+\epsilon\left(-4B_{0}m_{t}^{2}-2B_{1}\left(2m_{t}^{2}+t_{1}\right)\right)\biggr\},\label{eq:fs1_12}\end{eqnarray}
where the arguments of the scalar function and tensor coefficients
are $\left(\left(p_{g}-p_{t}\right)^{2},m_{t}^{2},0\right)=\left(t,m_{t}^{2},0\right)$. 
\item Bubble correction $S_{2}$:\begin{eqnarray}
f_{4}^{S_{4}} & = & \frac{1}{s}\left\{ 4B_{1}-4B_{1}\epsilon\right\} ,\\
f_{5}^{S_{4}} & = & \frac{1}{s}\left\{ 4B_{1}-4B_{1}\epsilon\right\} ,\\
f_{11}^{S_{4}} & = & \frac{1}{s}\left\{ 4B_{1}\epsilon-4B_{1}\right\} ,\\
f_{12}^{S_{4}} & = & \frac{1}{s}\left\{ 2B_{1}-2B_{1}\epsilon\right\} ,\end{eqnarray}
where the arguments of the scalar function and tensor coefficients
are $\left(\left(p_{g}+p_{b}\right)^{2},0,0\right)=\left(s,0,0\right)$. 
\end{itemize}

\section{Divergences of the form factors\label{sec:Divergences-of-formfactor}}

In this section we list out the divergent pieces of the form factors
and and we further distinguish the UV divergence and the IR divergence.
The former is written inside the square brackets, i.e. $\left[\cdots\right]$.
\\
(1) The triangle loop $V_{1}$ only gives rise to the UV divergences,\begin{equation}
f_{6}^{V_{1}}=2f_{12}^{V_{1}}=\frac{2}{t_{1}}\left[\frac{1}{\epsilon_{UV}}-2\right].\end{equation}
(2) The triangle loop $V_{2}$ exhibits both the UV and IR divergences,\begin{eqnarray}
f_{6}^{V_{2}} & : & \frac{1}{t_{1}}\left\{ \left[\frac{6}{\epsilon_{UV}}-4\right]-\frac{1}{\epsilon_{IR}^{2}}-\frac{4}{\epsilon_{IR}}-\frac{2}{\epsilon_{IR}}\ln\frac{-t_{1}}{m_{t}^{2}}\right\} ,\\
f_{12}^{V_{2}} & : & \frac{1}{t_{1}}\left\{ \left[\frac{3}{\epsilon_{UV}}-2\right]-\frac{1}{\epsilon_{IR}^{2}}-\frac{1}{\epsilon_{IR}}-\frac{2}{\epsilon_{IR}}\ln\frac{-t_{1}}{m_{t}^{2}}\right\} .\end{eqnarray}
(3) The triangle loop $V_{3}$ gives rise to both the UV and IR divergences,\begin{eqnarray}
f_{3}^{V_{3}} & : & \frac{1}{t_{1}}\left\{ -\frac{4}{\epsilon_{IR}}\frac{m_{t}^{2}-t}{m_{W}^{2}-t}\left(-1+\frac{m_{t}^{2}-m_{W}^{2}}{m_{W}^{2}-t}\ln\frac{-t_{1}}{m_{t}^{2}-m_{W}^{2}}\right)\right\} ,\\
f_{6}^{V_{3}} & : & \frac{1}{t_{1}}\left\{ \left[\frac{2}{\epsilon_{UV}}-4\right]+\frac{4}{\epsilon_{IR}}\left(-1+\frac{m_{t}^{2}-m_{W}^{2}}{m_{W}^{2}-t}\ln\frac{-t_{1}}{m_{t}^{2}-m_{W}^{2}}\right)\right\} ,\\
f_{12}^{V_{3}} & : & \frac{1}{t_{1}}\left\{ \left[\frac{1}{\epsilon_{UV}}-2\right]+\frac{2}{\epsilon_{IR}}\left(-1+\frac{m_{t}^{2}-m_{W}^{2}}{m_{W}^{2}-t}\ln\frac{-t_{1}}{m_{t}^{2}-m_{W}^{2}}\right)\right\} .\end{eqnarray}
(4) The triangle loop $V_{1}^{\prime}$ give rise to both the UV and
IR divergences,\begin{eqnarray}
f_{3}^{V_{1}^{\prime}} & : & \frac{1}{s}\left\{ \left[-\frac{2}{\epsilon_{UV}}+4\right]+\frac{4}{\epsilon_{IR}}\right\} ,\\
f_{4}^{V_{1}^{\prime}} & : & \frac{1}{s}\left[\frac{2}{\epsilon_{UV}}-4\right],\\
f_{5}^{V_{1}^{\prime}} & : & \frac{1}{s}\left\{ \left[\frac{2}{\epsilon_{UV}}-4\right]-\frac{4}{\epsilon_{IR}}\right\} ,\\
f_{11}^{V_{1}^{\prime}} & : & \frac{1}{s}\left\{ \left[-\frac{2}{\epsilon_{UV}}+4\right]+\frac{4}{\epsilon_{IR}}\right\} ,\\
f_{12}^{V_{1}^{\prime}} & : & \frac{1}{s}\left\{ \left[\frac{1}{\epsilon_{UV}}-2\right]-\frac{2}{\epsilon_{IR}}\right\} .\end{eqnarray}
(5) The triangle loop $V_{2}^{\prime}$ gives rise to both the UV
and IR divergences,\begin{eqnarray}
f_{3}^{V_{2}^{\prime}} & : & \frac{1}{s}\left\{ \left[-\frac{6}{\epsilon_{UV}}+4\right]+\frac{4}{\epsilon_{IR}^{2}}+\frac{6}{\epsilon_{IR}}-\frac{4}{\epsilon_{IR}}\ln\frac{-u-t_{1}}{m_{t}^{2}}\right\} ,\\
f_{4}^{V_{2}^{\prime}} & : & \frac{1}{s}\left\{ \left[\frac{6}{\epsilon_{UV}}-4\right]-\frac{2}{\epsilon_{IR}^{2}}-\frac{6}{\epsilon_{IR}}+\frac{2}{\epsilon_{IR}}\ln\frac{-u-t_{1}}{m_{t}^{2}}\right\} ,\\
f_{5}^{V_{2}^{\prime}} & : & \frac{1}{s}\left\{ \left[\frac{6}{\epsilon_{UV}}-4\right]-\frac{4}{\epsilon_{IR}^{2}}-\frac{6}{\epsilon_{IR}}+\frac{4}{\epsilon_{IR}}\ln\frac{-u-t_{1}}{m_{t}^{2}}\right\} ,\\
f_{11}^{V_{2}^{\prime}} & : & \frac{1}{s}\left\{ \left[-\frac{6}{\epsilon_{UV}}+4\right]+\frac{4}{\epsilon_{IR}^{2}}+\frac{6}{\epsilon_{IR}}-\frac{4}{\epsilon_{IR}}\ln\frac{-u-t_{1}}{m_{t}^{2}}\right\} ,\\
f_{12}^{V_{2}^{\prime}} & : & \frac{1}{s}\left\{ \left[\frac{3}{\epsilon_{UV}}-2\right]-\frac{2}{\epsilon_{IR}^{2}}-\frac{3}{\epsilon_{IR}}+\frac{2}{\epsilon_{IR}}\ln\frac{-u-t_{1}}{m_{t}^{2}}\right\} .\end{eqnarray}
(6) The triangle loop $V_{3}^{\prime}$ contains only the UV divergences:\begin{equation}
-f_{3}^{V_{3}^{\prime}}=f_{4}^{V_{3}^{\prime}}=f_{5}^{V_{3}^{\prime}}=-f_{11}^{V_{3}^{\prime}}=2f_{12}^{V_{3}^{\prime}}=\frac{2}{s}\left[\frac{1}{\epsilon_{UV}}-2\right].\end{equation}
(7) The box loops $B_{1,2,3}$ only contain the IR divergences. $B_{1}$
gives rise to the following IR divergences:\begin{eqnarray}
f_{3}^{B_{1}} & : & -\frac{1}{\epsilon_{IR}}\frac{4}{t-m_{W}^{2}}\left\{ 1+\frac{m_{t}^{2}-m_{W}^{2}}{t-m_{W}^{2}}\ln\frac{-t_{1}}{m_{t}^{2}-m_{W}^{2}}\right\} ,\\
f_{6}^{B_{1}} & : & \frac{1}{t_{1}}\left\{ -\frac{2}{\epsilon_{IR}^{2}}+\frac{4}{\epsilon_{IR}}\ln\frac{m_{t}^{2}-m_{W}^{2}-u}{m_{t}^{2}}+\frac{4}{\epsilon_{IR}}\frac{m_{t}^{2}-m_{W}^{2}}{t-m_{W}^{2}}\ln\frac{-t_{1}}{m_{t}^{2}-m_{W}^{2}}\right\} ,\\
f_{12}^{B_{1}} & : & \frac{1}{2}f_{6}^{B_{1}}.\end{eqnarray}
(8) $B_{2}$ gives rise to the following divergences:\begin{eqnarray}
f_{3}^{B_{2}}=-f_{5}^{B_{2}}=f_{11}^{B_{2}}=-2f_{12}^{B_{2}} & : & \frac{1}{s}\left\{ \frac{2}{\epsilon_{IR}^{2}}-\frac{4}{\epsilon_{IR}}\ln\frac{m_{t}^{2}-m_{W}^{2}-u}{m_{t}^{2}}\right\} ,\\
f_{4}^{B_{2}} & : & \frac{1}{s}\left\{ \frac{-2}{\epsilon_{IR}^{2}}+\frac{4}{\epsilon_{IR}}\ln\frac{m_{t}^{2}-m_{W}^{2}-u}{m_{t}^{2}}-\frac{4}{\epsilon_{IR}}\right\} .\end{eqnarray}
(9) $B_{3}$ gives rise to the following divergences:\begin{eqnarray}
f_{3}^{B_{3}} & : & \frac{1}{s}\left\{ \frac{4}{\epsilon_{IR}^{2}}+\frac{2}{\epsilon_{IR}}-\frac{4}{\epsilon_{IR}}\ln\frac{-t_{1}}{m_{t}^{2}}-\frac{4}{\epsilon_{IR}}\frac{m_{t}^{2}-m_{W}^{2}}{t-m_{W}^{2}}\ln\frac{-t_{1}}{m_{t}^{2}-m_{W}^{2}}\right\} \nonumber \\
 &  & -\frac{4}{\epsilon_{IR}\left(t-m_{W}^{2}\right)},\\
f_{4}^{B_{3}} & : & \frac{1}{s}\left\{ -\frac{4}{\epsilon_{IR}^{2}}-\frac{2}{\epsilon_{IR}}+\frac{4}{\epsilon_{IR}}\ln\frac{-t_{1}}{m_{t}^{2}}+\frac{2}{\epsilon_{IR}}\ln\frac{s}{m_{t}^{2}}\right\} ,\\
f_{5}^{B_{3}} & : & \frac{1}{s}\left\{ -\frac{2}{\epsilon_{IR}^{2}}-\frac{2}{\epsilon_{IR}}+\frac{4}{\epsilon_{IR}}\ln\frac{-t_{1}}{m_{t}^{2}}\right\} ,\\
f_{6}^{B_{3}} & : & \frac{1}{t_{1}}\left\{ -\frac{5}{\epsilon_{IR}^{2}}+\frac{4}{\epsilon_{IR}}\ln\frac{s}{m_{t}^{2}}+\frac{4}{\epsilon_{IR}}\frac{m_{t}^{2}-m_{W}^{2}}{t-m_{W}^{2}}\ln\frac{-t_{1}}{m_{t}^{2}-m_{W}^{2}}+\frac{2}{\epsilon_{IR}}\ln\frac{-t_{1}}{m_{t}^{2}}\right\} ,\\
f_{11}^{B_{3}} & : & \frac{1}{s}\left\{ \frac{2}{\epsilon_{IR}^{2}}+\frac{2}{\epsilon_{IR}}-\frac{4}{\epsilon_{IR}}\ln\frac{-t_{1}}{m_{t}^{2}}\right\} ,\\
f_{12}^{B_{3}} & : & \frac{1}{s}\left\{ -\frac{1}{\epsilon_{IR}^{2}}-\frac{1}{\epsilon_{IR}}+\frac{2}{\epsilon_{IR}}\ln\frac{-t_{1}}{m_{t}^{2}}\right\} \nonumber \\
 &  & +\frac{1}{t_{1}}\left\{ -\frac{2}{\epsilon_{IR}^{2}}-\frac{1}{\epsilon_{IR}}+\frac{2}{\epsilon_{IR}}\ln\frac{s}{m_{t}^{2}}+\frac{2}{\epsilon_{IR}}\frac{m_{t}^{2}-m_{W}^{2}}{t-m_{W}^{2}}\ln\frac{-t_{1}}{m_{t}^{2}-m_{W}^{2}}\right\} .\end{eqnarray}
(10) The bubble loops $S_{1,2}$ only contain the UV divergences.
$S_{1}$ gives rise to the following divergences: \begin{eqnarray}
f_{2}^{S_{1}} & : & \frac{m_{t}}{t_{1}}\left[\frac{3}{\epsilon_{UV}}-1\right],\\
f_{6}^{S_{1}} & : & \frac{2}{t_{1}}\left[\frac{1}{\epsilon_{UV}}\frac{6m_{t}^{2}}{t_{1}}-\frac{1}{\epsilon_{UV}}-\frac{2m_{t}^{2}}{t_{1}}+1\right],\\
f_{12}^{S_{1}} & : & \frac{1}{t_{1}}\left[\frac{1}{\epsilon_{UV}}\frac{6m_{t}^{2}}{t_{1}}-\frac{1}{\epsilon_{UV}}-\frac{2m_{t}^{2}}{t_{1}}+1\right],\end{eqnarray}
while $S_{2}$ gives rise to the following divergences:\begin{equation}
f_{3}^{S_{2}}=-f_{4}^{S_{2}}=-f_{5}^{S_{2}}=f_{11}^{S_{2}}=-2f_{12}^{S_{2}}=\frac{2}{s}\left[\frac{1}{\epsilon_{UV}}-1\right].\end{equation}

\bibliographystyle{apsrev}
\clearpage\addcontentsline{toc}{chapter}{\bibname}\bibliography{reference}

\end{document}